# Saving temporary exhibitions in virtual environments: the Digital Renaissance of Ulisse Aldrovandi – acquisition and digitisation of cultural heritage objects

Balzani Roberto[a], Barzaghi Sebastian[b], Bitelli Gabriele[c], Bonifazi Federica[d], Bordignon Alice[d], Cipriani Luca[e], Colitti Simona[e], Collina Federica[b]*, Daquino Marilena[d], Fabbri Francesca[b], Fanini Bruno[f], Fantini Filippo[e], Ferdani Daniele[f], Fiorini Giulia[c,g], Formia Elena[e], Forte Anna[e], Giacomini Federica[b], Girelli Valentina Alena[c], Gualandi Bianca[d,i], Heibi Ivan[d], Iannucci Alessandro[b], Manganelli Del Fà Rachele[h], Massari Arcangelo[d], Moretti Arianna[d], Peroni Silvio[d], Pescarin Sofia[h], Renda Giulia[d], Ronchi Diego[f], Sullini Mattia[e], Tini Maria Alessandra[c], Tomasi Francesca[d], Travaglini Laura[d], Vittuari Luca[c]

[a] *Department of History and Cultures, University of Bologna, Piazza S. Giovanni in Monte, 2, 40124 Bologna (BO), Italy*
[b] *Department of Cultural Heritage, University of Bologna, Via degli Ariani, 1, 48121 Ravenna (RA), Italy*
[c] *Department of Civil, Chemical, Environmental, and Materials Engineering, University of Bologna, Viale del Risorgimento, 2, 40136 Bologna (BO), Italy*
[d] *Department of Classical Philology and Italian Studies, University of Bologna, Via Zamboni, 32, 40126 Bologna (BO), Italy*
[e] *Department of Architecture, University of Bologna, Viale del Risorgimento, 2, 40136 Bologna (BO), Italy*
[f] *Digital Heritage Innovation Lab, Institute of Heritage Science, National Research Council, Via Salaria km 29,300, 00010 Montelibretti, Rome (RM), Italy*
[g] *Department of Classics, Sapienza University of Rome, Piazzale Aldo Moro 5, 00185, Rome, Italy*
[h] *Digital Heritage Innovation Lab, Institute of Heritage Science, National Research Council, Via Madonna del Piano, 10, 50019 Sesto Fiorentino (FI), Florence, Italy*
[i] *Research Services Coordination Unit, Research Division, University of Bologna, Via Zamboni, 32, 40126 Bologna (BO), Italy*

*Corresponding author: Federica Collina, Department of Cultural Heritage, University of Bologna, Via degli Ariani, 1, 48121 Ravenna RA, Italy, federica.collina5@unibo.it

Declarations of interest: none

ABSTRACT

As per the objectives of Project CHANGES, particularly its thematic sub-project on the use of virtual technologies for museums and art collections, our goal was to obtain a digital twin of the temporary exhibition on Ulisse Aldrovandi called "The Other Renaissance", and make it accessible to users online. After a preliminary study of the exhibition, focussing on acquisition constraints and related solutions, we proceeded with the digital twin creation by acquiring, processing, modelling, optimising, exporting, and metadating the exhibition. We made hybrid use of two acquisition techniques to create new digital cultural heritage objects and environments, and we used open technologies, formats, and protocols to make available the final digital product. Here, we describe the process of collecting and curating bibliographical exhibition (meta)data and the beginning of the digital twin creation to foster its findability, accessibility, interoperability, and reusability. The creation of the digital twin is currently ongoing.

# 1. Introduction

Several international policies support the focus on the universal use of digital data. Expressly, on Cultural Heritage (CH), both UNESCO and the EU have already provided clear guidelines for increasing the digitisation of heritage, considering aspects related to cataloguing systems as well as different ways for accessing cultural heritage objects and sites (e.g. *in situ* or decentralised). However, in the Italian context, expert users often perceive CH as composed of tangible and intangible artefacts. In particular, Italian legislation and established and shared practices focused primarily on using material objects. To comply with current European policies and guidelines on CH, we need to go beyond this configuration and make its digital enhancement a permanent and widespread practice in museums and art collections, aiming at (a) increasing the knowledge, curation and management of artefacts in all forms, (b) expanding the involvement of the general public and communication techniques, (c) improving and making more sustainable the exhibition potential, and (d) including crucial social functions such as accessibility, inclusiveness, critical thinking, participation, enjoyment and sustainability.

The Project CHANGES ("Cultural Heritage Active Innovation For Next-Gen Sustainable Society") is an EU-funded national project that started in December 2022 and involves several Italian actors – 11 universities, 4 research institutions, 3 schools for advanced studies, 6 companies, and 1 centre of excellence – aiming at promoting and preserving Italian cultural heritage in compliance with the aforementioned European policies. The authors of this work are involved in Spoke 4, a thematic sub-project specifically dedicated to the use of virtual technologies for the promotion, preservation, exploitation and enhancement of CH in museums and art collections.

One of the main goals of Spoke 4 is the identification of potential technologies, methodologies and solutions that could solve the main questions and priorities cultural institutions (and national collections) are experiencing today. Part of our effort, therefore, is also to explore and involve national cultural institutions – which include natural history and scientific museums, widespread art galleries, sites museums with (in)tangible heritage and landscapes, historical palaces, demo-ethnic-anthropological museums and also museums with large collections and high-tech approach – in a dialogue meant to specify these questions and define together a priority and find solutions, prototyping examples and applications that they can use. Indeed, we have identified several *core* case studies, involving different cultural institutions as representative of the different museum contexts in Italy, to define best practices that can be further adapted and reused in institutions and contexts sharing similar characteristics.

One of the first goals of Spoke 4 has been to identify a set of guidelines that can support the researchers and other members of the Spoke to set up appropriate cultural heritage acquisition processes. Thus, before developing final solutions for the case studies, we have identified a scenario that could serve as a common experimental ground for a multidisciplinary group and as a baseline to define some approaches and methods relating to the acquisition, processing, optimisation, metadata inclusion and online publication of 3D assets. For this, we selected a *temporary* exhibition (ended on 28th May 2023) containing a large set of different small/medium objects to be acquired and entitled "The Other Renaissance: Ulisse Aldrovandi and the Wonders of the World".

The overall goal was to obtain a *digital version of the experience at the exhibition*, starting from its *digital twin*, connected to the digital asset of the different items (3D and multimedia) of the collections, organised and accessible online by users, using various devices (from home computers, smartphones, to tablets and VR headsets). In particular, we have identified four main research questions (RQ1-RQ4) to be addressed:

1. How to preserve and make accessible temporary exhibitions and specifically the potential experience a visitor can have?
2. What is the impact of environmental, temporal and contextual constraints (i.e. objects in glass cases, mandatory short time of acquisition, etc.) in the final digital result and how to keep track of this impact in the final stored or even visualised object?
3. How to handle, acquire and process objects made of critical and peculiar material (such as grass, dimension, fur, and metal)?
4. How to realise the entire data flow, from the management of the acquisition process to digital data and metadata on-line access for experts, from metadata storage to interactive experiences, such as virtual exhibitions/virtual museums or digital libraries?

5. How can we ensure the proposed approach would be affordable (in time and cost), also to fulfil the usual limited budget of temporary exhibitions?

In this article, we introduce all the methods and processes adopted to deal with the initial creation phase of the digital twin of the temporary exhibition, which allowed us to reflect on and address the aforementioned research questions and to acquire and digitise the cultural heritage objects included in the exhibition. The finalisation of the digital twin of the temporary exhibition is still ongoing, and it will be presented in detail in future works.

The rest of the paper is structured as follows. In Section 2, we introduce the context of our work, the temporary exhibition we worked on and other concepts related to virtual exhibitions and digital twins in CH. In Section 3, we detail some of the most relevant works related to our research. In Section 4, we present the materials and strategies developed to approach the digitisation of the exhibition. In Section 5, we focus on the most relevant aspects related to the acquisition and processing of the objects of the temporary exhibition. In Section 6, we introduce the necessary steps for creating and publishing the virtual exhibition on the Web. Finally, in Section 7, we discuss the main results obtained so far in our research and conclude the article by sketching out some future works.

## 2. The context

In this section, we introduce some contextual information needed for the reader to understand and appreciate the context behind our work. In particular, we describe the temporary exhibition we have worked on, presenting relevant literature discussing the technologies, standards and formats to create a digital replica (i.e. a *digital twin*) of visitors' experience at the exhibition.

### 2.1. The exhibition "The Other Renaissance"

The exhibition "The Other Renaissance: Ulisse Aldrovandi and the wonders of the world" (https://site.unibo.it/aldrovandi500/en/mostra-l-altro-rinascimento) was held in Poggi Palace Museum (Via Zamboni 33, Bologna, Italy) between December 2022 and May 2023. The Museum is located inside Palazzo Poggi, a palace built in the 16th century to be the home of the Poggi family. In 1711, it was acquired by the Bolognese Senate, becoming the Institute of Science and the Arts there (Rosati, 2020). With the Napoleonic reforms, the University was moved there, and the original historical collections were dismembered and redistributed, according to the various research fields, among specialised university institutes and city museums. At the end of the 20th century, thanks to a cultural reorganisation and restoration project, the Aldrovandi collection was relocated to this palace (Latini, 2005).

Aldrovandi's research belongs to the phase of 'archaic globalisation' (Bayly, 2004), defined by the idea of universal monarchy, the expansive thrust of the great religions and the humoral or moral conception of the health of the body (Aldrovandi was also a physician). It was a pre-colonial phase, including scholars from geographical areas not involved in constructing empires. Aldrovandi's view, although conditioned by the prejudices of his time, was guided above all by *curiositas* (Pomian, 2023), trying to document the extraordinary variety of the natural world. The exhibition organisers considered this aspect critical and proposed an approach that enabled them to recover evidence of an alternative way of looking at relations between mankind, space and nature, based primarily on a pure desire for knowledge (Carrada, 2022). Aldrovandi's *theatre of nature* in Bologna represented a *unicum* in 16th-century Europe (Pomian, 2020). The creation of a digital manifestation of the exhibition was therefore also conceived to broaden the audience of people interested in this ancient message of peaceful knowledge, which, by the way, represents one of the cornerstones of the University identity worldwide.

The idea that guided this experience was to introduce Aldrovandi as a forerunner of modern scholarly and scientific communication. The detailed description of the process that enabled him to introduce objects in illustrated volumes, which emphasised the close relationship between art and science (Olmi and Simoni, 2018), made it possible to show the effort to represent the natural real together with the residual interferences of the imaginary, which is a legacy of the Middle Ages. The digital twin of the exhibition thus represents a further step in the dissemination process: it is the most modern development of a 500-year-old intuition. For this reason, too, the Aldrovandi case has a special scientific significance and moral message.

The exhibition was conceived to guide visitors to discover what was called the "reawakening of natural sciences", a moment of the Renaissance, in which the naturalist Ulisse Aldrovandi in Bologna and a small group of doctors, pharmacists and naturalists (who first studied animals and plants from life, rather than just in books), took the first steps towards science as we know it today (Rebellato, 2018). The exhibition was developed in six rooms of the Palace (the first two shown in Fig. 1) as a tour of a collection of more than 200 objects, most belonging to Aldrovandi and preserved by the University of Bologna. This collection of 'jewels', most of which have never been exhibited before, included objects and works of art from other Italian museums, but also multimedia material included in the itinerary (Haxhiraj, 2016). This unique exhibition received a high appreciation from the public (more than 30,000 visitors), leading the curators to postpone the closing of such extraordinary objects and the storytelling that accompanied the visitors in this incredible discovery of how science was born. As it was a temporary exhibition, its conclusion was perceived as a lost occasion for the many potential other citizens and scholars who could have been interested in revisiting it.

This open issue, common to the many temporary exhibitions around the world, has led us to use it as an exceptionally appropriate case to focus on this priority and search for potential solutions supported by virtual technologies.

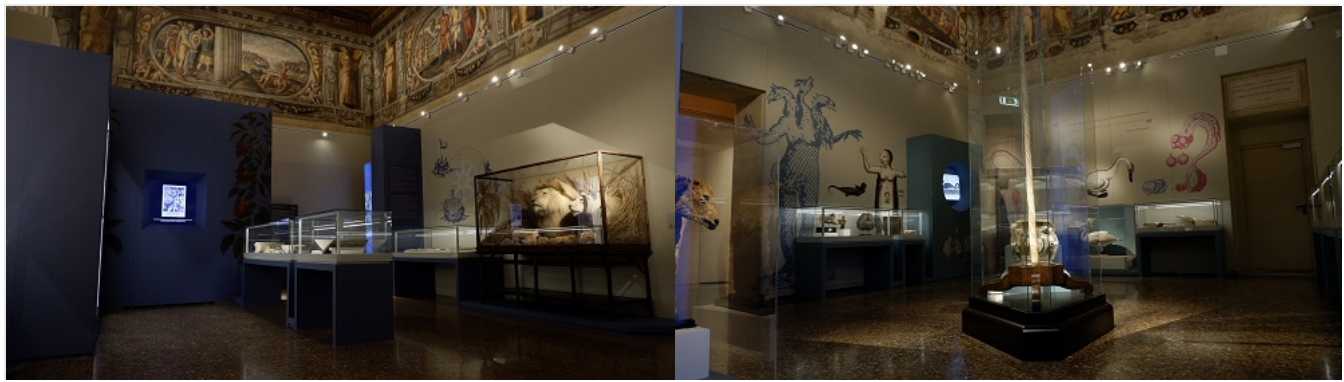

Fig. 1. Two pictures of the first two rooms of the exhibition "The Other Renaissance: Ulisse Aldrovandi and the wonders of the world" - source: https://site.unibo.it/aldrovandi500/en/mostra-l-altro-rinascimento.

*2.2. Towards a virtual exhibition: 3D research data and Digital Libraries*

To maximise the re-adoption of our workflow for the creation of a virtual exhibition in other contexts (internally or externally to Spoke 4), we used as many open tools and software as possible, performing a selection of open-source software spanning, at least in principle, all the steps of our workflow. However, when appropriate, we used proprietary software making sure to use standard formats whenever possible – e.g., for 3D models (glTF, GLB, obj, mtl, png, tiff, jpg, e57), images (tiff, png, jpg, raw), video (mp4, mov), and audio (mp3) – as not to be bound to any proprietary applications and in line with the prescription of the Spoke 4 Data Management Plan (Gualandi and Peroni, 2023). In particular, regarding 3D formats, we pushed the adoption of glTF, an interoperable open standard targeting interactive Web3D applications to guarantee high interoperability with existing 3D platforms/services, re-use and integration of licensing information inside the format. In terms of licences used for publishing the various research data, we tried to be as open as possible, following the EU directives and the constraints imposed by Italian law. Thus, the current set of licences adopted for the majority of gathered data (e.g. descriptive metadata and 3D objects depicting existing real-world cultural heritage objects) varied from very permissive ones (i.e. Creative Commons Zero – https://creativecommons.org/publicdomain/zero/1.0/legalcode) to more restrictive ones (i.e. Creative Commons Attribution Non-Commercial – https://creativecommons.org/licenses/by-nc/4.0/legalcode). Instead, for the objects in the exhibition that are not directly owned and curated by the University of Bologna, we agreed on specific licences according to the will of the source institutions.

Our work takes into account the European and national context (Bucciero et al., 2022). In particular the recent calls and projects related to the Infrastructures for the Heritage Sciences (E-RIHS: https://www.e-rihs.eu/), the challenging goal set by the European Commission to digitise the entire Cultural Heritage at risk within 2030 (and 40% of it before 2025) through specific calls aimed at creating a "dataspace" for the cultural patrimony (source: Europeana conference 2022), and the launch of the European Collaborative Cloud for Cultural Heritage (ECCCH). The Italian Ministry of Cultural Heritage (MIC) is also actively working in this direction, with the publication of the National Plan for the Digitisation of Cultural Heritage (PND) in 2022 by the Istituto Centrale per la Digitalizzazione del Patrimonio Culturale – Digital Library. The plan promotes and organises the process of digital transformation in the period 2022-26 and contains several guidelines that provide a robust technical and methodological framework for planning and implementing activities related to the digitisation of cultural heritage (PND, 2022). In detail, the Guidelines for Digitisation define approaches and procedures for the creation, metadata processing and archiving of digital objects, including 2D, 3D, audio and video digital resources, in line with the so-called FAIR management of research data.

FAIR principles state that data should be Findable, Accessible, Interoperable and Reusable, leaving it to the different research communities to figure out how exactly these "deceptively simple" principles can be put into practice in the respective disciplines (Knazook et al., 2023b). Efforts have recently been made to describe how FAIR principles can be interpreted for and applied to the cultural heritage sector (Koster and Woutersen-Windhouwer, 2018), where digital collections are ever-evolving research data that need to be amenable to computational use (Padilla et al., 2019). Some key criticalities, such as the inter woveness of data and metadata (Koster and Woutersen-Windhouwer, 2018; Tóth-Czifra, 2019) or the need to further qualify cultural heritage data re-use (Barbuti, 2021) have also been highlighted.

However, little attention is given to 3D models as research data, perhaps because their adoption in the cultural heritage sector is currently low (Knazook et al., 2023b). A recent report simply suggests that "the purpose and value of 3D imaging [...] will probably be very different from traditional digital images" and "may require different policies, as well as technologies, to make them accessible and preservable" (Knazook et al., 2023a).

*2.3 Digital Twins*

Since our overall goal is to let visitors and experts access the temporary exhibition after its closing, we have started with the creation of a Digital Twin, a necessary step to develop more complex VR experiences, such as Virtual Museums (Pescarin, 2014).

The concept of Digital Twins originated from established technologies like 3D modelling, simulation, and prototyping that have enabled tighter integration between physical and virtual worlds (Errandonea et al., 2020; Qi et al., 2021). Digital Twins create virtual counterparts of physical entities to replicate properties, states, and behaviours for assessment, optimisation, prediction, and simulation (Semeraro et al., 2021). The term was first coined by Grieves and Vickers (2017) to define a digital informational representation tethered to the physical counterpart via continuous data flows (Errandonea et al., 2020; Qi et al., 2021). The concept underwent further development through the publication of the Gemini Principles report (Bolton et al., 2018), which established fundamental guidelines to foster consensus on information management practices in the UK's built environment sector. According to these principles, digital twins are "accurate digital representations of physical assets, processes, or systems in the built or natural environment", whose significance lies in their ability to provide valuable insights for informed decision-making. These principles underscore the necessity for digital twins to possess appropriate levels of detail tailored to their specific objectives. Furthermore, they illuminate the diverse roles of digital twins (encompassing functions such as prospective scenario planning, real-time monitoring and control, and historical analysis) and their versatility in terms of spatial and temporal scopes.

The complexity surrounding Digital Twins has promoted the proposal of conceptual dimensions or characteristics to categorise them according to different interpretations (van der Valk et al., 2020; Qi et al., 2021). Digital Twins commonly involve virtual-physical pairing, data connections, and simulation capabilities (VanDerHorn and Mahadevan, 2021). Key elements in Digital Twin definitions include physical/virtual entities, environments, connections, metrics, rates, and more (Jones et al., 2020). Interaction, bidirectional data exchange, update, synchronisation, and simulation are also pivotal concepts, spanning both data and algorithms that make up the Digital Twin.

Further classifications have been suggested based on the integration level between physical objects and digital replicas. These include:

- *Digital Model* – a digital representation of an existing or planned physical object with no automatic data flow between the physical object and its digital copy;
- *Digital Shadow* – a Digital Model which includes a unidirectional automated data flow from the physical object to its copy;
- *Digital Twin* – a Digital Shadow that includes a bidirectional data flow (Kritzinger et al., 2018).

However, some argue that certain Digital Twin requirements are too restrictive and thus propose a generalisation that converges on the essential elements while avoiding use-case-specific terminology (VanDerHorn and Mahadevan, 2021). Others recognise that precise manifestations vary across contexts and applications based on specific uses and methods (Errandonea et al., 2020; Niccolucci et al., 2023). Thus, amidst foundational agreement, the Digital Twin concept can be flexible enough in realisation according to the different domains, objectives and approaches.

This applies especially well in the cultural heritage domain. Notably, the existing literature only scratches the surface in examining Digital Twins, specifically in cultural heritage, primarily in relation to Building Information Models (BIMs) (Azhar et al., 2020) and Heritage BIMs (HBIM) (López et al., 2018) as a starting point for implementing Digital Twin capabilities applied to the cultural heritage domain (Bitelli et al., 2017; Jouan and Hallot, 2020; Shahzad et al., 2022). However, fully implementing a Digital Twin for CH faces significant obstacles since many prescribed Digital Twin requirements may not suit the unique nature of CH (Gabellone, 2022), which often involves intangible aspects, as well as dynamic and complex changes over time due to various factors, that could even cause the physical object to no longer exist.

Yet, the potential exists to re-conceptualize Digital Twins for CH using frameworks like HBIM. Niccolucci et al. (2023) suggest decoupling the dimension of data representation from that of data exchange, stating that, in the cultural heritage context, features like continuous data exchange and bi-directionality, while possible, should not be requirements since heritage may be immaterial or temporary. This foundation can evolve dynamically to serve specific uses, forming an accurate, versatile "model of knowledge" closer to the notion of a Digital Model that can be extended to eventually become a fully-fledged Digital Twin.

*2.4 The Digital Renaissance of Ulisse Aldrovandi*

As anticipated, the creation of a digital twin of the exhibition "The Other Renaissance: Ulisse Aldrovandi and the Wonders of the World" was, in the first instance, functional to preserve an event that ceased to exist at the end of May 2023. Here, the idea is to enable users to continue accessing the exhibition from their homes by entering the exhibition spaces via a browser, using their own devices and following the original flow of the narrative created by the curators to bring the users around the masterpieces exposed in the rooms of Palazzo Poggi. However, the digital exhibition already provides at least an additional level of interaction which was impossible in the physical environment: to go beyond the purely static observation of the objects behind the display cases and take such objects with *digital* hands, zooming and rotating them as the users prefer. Thus, the digital approach allows users to explore the breadth and depth of Aldrovandi's work, from his study of natural history to his correspondence and other related materials, providing insight into the naturalist's working environment and the materials he used in his research.

In addition, by leveraging a combination of 3D, Linked Open Data (Bizer et al., 2009), Knowledge Graphs (Hogan et al., 2021), and Web and Semantic Web technologies (Berners-Lee et al., 2001), we can expose semantic relations existing between the exhibition and other external objects, as well as the exhibition itself, to enable users to explore it on their terms and gain access to a broader cultural and historical context. This approach also provides an opportunity to create new narratives beyond the physical exhibition and serve as a starting point for new interactive experiences. For example, users can be invited to contribute with their content, e.g. via annotations on the digital objects included in the exhibition, to create a collaborative and participatory experience. This can foster a sense of community and engagement around the exhibition and provide a platform for users to share their perspectives and interpretations, thus fostering social cohesion (Pescarin et al., 2023a). Additionally, this approach allows for easier integration with other existing digital resources (Fiorini et al., 2022), further enriching it with additional information and making it easier for researchers and enthusiasts to discover, explore and reuse Aldrovandi's work. Finally, this approach provides invaluable support for the documentation of the exhibition, further enhancing its representation, accessibility, and reuse within the scientific community and the public.

## 3. State of the Art

### 3.1. 3D models published by cultural institutions

Several museums have been experimenting with making 3D data and multimedia available to the public (Bitelli et al., 2022). For example, as of July 2023, the Smithsonian has 2685 3D models (https://3d.si.edu/); 244 of them have been released under Creative Commons Zero (https://creativecommons.org/publicdomain/zero/1.0/legalcode), while re-use conditions, typically banning commercial exploitation, apply in all other cases. These models can be viewed on the museum's website, via a custom 3D viewer, or on the SketchFab platform (https://sketchfab.com/).

The British Museum also uses SketchFab to display its 277 models[1], mostly shared under a Creative Commons Attribution-NonCommercial-ShareAlike (https://creativecommons.org/licenses/by-nc-sa/4.0/legalcode) licence. A selection of these 3D models is also showcased on the museum's website (https://sketchfab.com/britishmuseum/models) by embedding the SketchFab 3D viewer.

The Naturhistorisches Museum Wien (NHMW) has a section titled 3D museum (https://www.nhm-wien.ac.at/en/museum_online/3d) that contains around 200 objects – including the "top 100 objects" in the museum – published under a Creative Commons Attribution-NonCommercial (https://creativecommons.org/licenses/by-nc/4.0/legalcode) license. These models can be best viewed on SketchFab, but its embed option has been used to display some of the models on the museum's website, too. Interestingly, the institution also has a Data Repository (http://datarepository.nhm-wien.ac.at/) run by the museum's publishing arm in collaboration with the NHMW Central Research Laboratories and IT department. This archive contains datasets, images and textual sources but not the aforementioned 3D models.

The Museum fur Naturkunde Berlin has also proactively invested in research data management infrastructures. Its Data portal (https://portal.museumfuernaturkunde.berlin/) contains over 75,000 media files, including images, audio files and around 25 3D models. These models are all published under a Creative Commons Attribution (https://creativecommons.org/licenses/by/4.0/legalcode) licence and can be viewed directly within the data portal. The museum's 3D lab has its collection of models on SketchFab (https://sketchfab.com/VisLab), but there does not seem to be any formal link to or from the museum's website or data repository.

### 3.2. Experiments in creating digital twins

Recent work demonstrates the growing use of digital twin technologies for cultural heritage applications, especially for conservation purposes. Apollonio et al. (2017) developed an innovative 3D web system to document and manage the restoration of the Neptune Fountain in Bologna. This pioneering digital twin integrated high-resolution 3D scans of the fountain with diagnostic data about its condition and restoration needs. By combining 3D visualisation and information management, the system enabled detailed recording and analysis of the fountain's state before, during, and after the restoration campaign. Girelli et al. (2019) provide further technical details on the 3D survey methodology used to capture the high-fidelity digital twin of the Neptune Fountain. Precise 3D scanning and modelling were crucial to creating an accurate digital replica supporting an information system for collecting and managing diagnostics data.

---

[1] https://sketchfab.com/britishmuseum/models

La Russa and Santagati (2020) illustrate an innovative application of digital twin principles for museum collection conservation and management. They developed a methodology combining AI, BIM, and IoT called Historical Sentient-Building Information System (HS-BIM) applied to the MuRa museum at Villa Zingali Tetto. The HS-BIM digital twin model perceives many internal and external data inputs from the historic building, including 3D scans, weather data, and microclimate analysis. By integrating these real-time sensor feeds with a virtual model of the building, the AI-enhanced digital twin can investigate complex relationships between variables to inform predictive strategies for optimising museum conservation. This novel use of bidirectional data flows between the physical building and virtual model for collection care highlights the potential of digital twins in cultural heritage for driving data-based decision-making, preventive conservation, and environmental management. The autonomous "sentient" capabilities demonstrate cutting-edge digital twin implementations that point to the future of smart, self-adapting museums.

Kong and Hucks (2023) proposed an innovative photogrammetry-based framework for creating digital twin models to monitor the deterioration of infrastructure over time. Their methodology involves creating two high-fidelity 3D models of a site using photogrammetry – one at an initial time point and another after a period of deterioration. By aligning the twin models and applying point cloud comparison algorithms, they can precisely identify and analyse geometric changes between inspections. This approach was demonstrated on the Taleyfac Spanish Bridge in Guam to quantify and visualise concrete cracking, spalling, and other deterioration.

Gros et al. (2023) pioneered a digital twin approach for precise virtual reconstruction after disasters, applied to the damaged arch of Notre Dame Cathedral. Using an iterative modelling methodology, they obtained a quantitative 3D model of the fallen arch starting from its fragmented remains. With over 73% accuracy, their digital twin enabled an analytical assessment of the arch's original state to support reconstruction. This showcases digital twins' capabilities for disaster recovery and restoration of heritage sites.

Gabellone (2022) employed IoT sensors, 3D reconstruction and interactive features to create a digital twin of the Gallipoli oil mill. This virtual model recreated the historic oil mill and allowed remote users to explore the site through guided tours, interacting with a live guide and engaging in multimedia educational content. A 3D Vista web application enhanced the experience with multi-user functionality for shared virtual visits.

In other cases, the focus is on representing rich contextual information characterising cultural heritage. Bevilacqua et al. (2022) developed an innovative "during-time Digital Twin" to reconstruct the historical transformations of the Duomo of Carpi Cathedral over centuries. Leveraging technologies like BIM, AR and VR, they created a 3D model illustrating the complete evolutionary timeline of the church. This novel application of digital twin principles aimed to embed the rich historic transformations directly into the virtual model, enabling analysis and visualisation of the site's architectural history. Their work demonstrates the potential of digital twins for predicting future changes and reconstructing and representing the complex past evolutions of heritage sites.

Similarly, Guzzetti et al. (2023) created a digital twin of the Ghirlanda of the Castello Sforzesco in Milano to collect all the evolutionary phases of the architectural asset. The 3D model encompasses the actual architectural structure, those that no longer exist. The digital twin has been equipped with architectural, documental, and textual data. The objective is to create a tool that facilitates the comprehension of structures and embeds in a unique corpus all the pieces of information related, otherwise fragmented. The structure has been acquired with three levels of geometric information, differentiated in no longer existing, reconstruction hypothesis or whether they present only a document collection, and modelled on BIM. The result is a collection of all the elements of the building where each element of the 3D can be updated.

Niccolucci et al. (2023) proposed the "Heritage Digital Twin" concept to encompass diverse knowledge of heritage assets in a semantic ontology. Moving beyond 3D models, they developed an ontology to represent multi-faceted information as structured linked data. Implementing this for Vasari's Last Supper and the Pafos historic gateway, they demonstrated encoding physical and intangible characteristics like materials, construction, provenance and more. This knowledge-based approach integrates digital twins with ontologies and linked data to enable sophisticated preservation and management of cultural heritage.

*3.3. Constraints on digitisation processes*

Temporary exhibitions, such as many museum exhibitions, present several problems from the point of view of digitisation (Gattet et al., 2015; Di Paola et al., 2022; Farella et al., 2022). The three main complexities encountered in the literature were related to time, space, and materials.

The first constraint, related to the *time* factor, was a direct consequence of the temporary nature of an exhibition, just as it may well be related to the short acquisition time a project may have. The short time available to complete acquisitions, therefore, forces one to choose not only the method that can guarantee the best results but also the most efficient way to save time and energy. Upstream, there is a tendency to avoid acquiring a certain type of object, and so surface/material, with an instrument whose problems one already knows about when acquiring it (Di Paola et al., 2022).

The second complexity involves *space*. The exhibition spaces, in general, are designed for visitation and certainly not for allowing the different research groups to acquire the objects in comfort and at the same time. Thus, in approaching acquisition, one has to deal with coordination among the various research groups and the lack of support, the scarcity of electrical outlets, suitable lighting conditions, and the few "manoeuvring spaces" available. Furthermore, due to their delicacy, weight, or display mode, many of the assets cannot be handled for acquisition. In these cases, objects must be digitised within their display cases, or it becomes necessary to change the acquisition settings to adapt them to their position (Bruno et al., 2010; Gonizzi Barsanti and Guidi, 2013). The lighting setup is another logistical aspect that needs careful planning during the acquisition phase. That is necessary because museum environments tend to be dimly lit, and achieving proper lighting for high-quality photography or a good scanner acquisition, requires thoughtful arrangement of lighting equipment (McPherron et al., 2009; Dall'Asta et al., 2016; Farella et al., 2022).

Finally, a further problem regarding acquisition concerns the *materials* of which some works are composed: several objects may present non-cooperative or hybrid materials, including specular components that could influence the optical response of the detection instruments, such as black, glossy or transparent surfaces (Bruno et al., 2010; Guidi et al., 2010). In the experience presented here, the photographic acquisition posed specific and additional challenges. First and foremost, the physical dimensions of the objects could exceed the depth of field achievable. Either to get images with a good sampling distance and given the operative distance from the subject in the range of 50-150 cm due to the tight spaces, the achievable depth of field for a standard circle of confusion is less than 1 cm, even for the shortest focal lengths. However, it is still usable up to 10 cm with a minimum loss of sharpness, a quantity still insufficient to include the whole object anyway. When these cases occurred, objects were acquired using the focus stacking technique. This approach consented to minimising the problem by shooting multiple images with different focus distances from each viewpoint (Santella and Milner, 2017). By doing so, the photogrammetric software is fed with images where different portions of the object were in focus even if the point of view was the same for all these subsets of images, leading to a more precise reconstruction and especially of the textures. Quality could be further improved by masking only the areas in focus within each image, or images can be pre-processed by merging them into one image with all parts in focus. Still, even without this operation, the quality gain is valuable.

Additionally, the same problem with DoF required special attention for objects characterised by prominent parts, especially if they presented finer details than the main body of the object. This has been the case with taxidermied fishes with long, thin fins, eventually with frayed edges, vase handles, or paws and ears of taxidermied mammals. All these parts would be out of focus if pointing at the body and then have been treated as if they were independent of the main object by dedicating specific subsets of images centred and focused on each of them. This approach provides the software with enough footage to reconstruct these parts with completeness and a level of detail on par with the rest of the object.

In some cases, objects were not movable or, due to their fragility, could be handled with limitations and only under the oversight of a supervisor. Such situations posed further limiting factors regarding lighting conditions control. Hence, some ploys had to be found to use the existing light sources on site while limiting their main drawbacks: sub-optimal placement, illuminance (hard lights), and light temperature. One way to address this aspect is by repositioning the object whenever possible, allowing the handling constraints to obtain the lowest contrast between directly lit and shadowed surfaces and creating enough space around the object for the operator to move. A low contrast reduces the risk of having dark or bright areas out of the optimal range of the camera, thus improving the effectiveness of 3D reconstruction. At the same time, the freedom of movement around the object enables collecting as many viewpoints as possible to enhance the 3D reconstruction and make it more complete and affordable for the software.

Regarding illuminance, the scope is to avoid as much as possible hard highlights on the object surface when it has a reflective component hence it is not comparable to a Lambertian material because hard highlights may cause misalignments of cameras or holes in the 3D model. This can be obtained either by avoiding direct light from concentrated sources such as spotlights and preferring indirect light instead or by screening the objects with translucent materials such as sheets that would distribute the light amount emitted on a wider surface with a great reduction of the illuminance, even at the cost of a slight reduction in the amount of light that actually would hit the object. This consequence does not impact the data quality since shutter speed is not a concern unless there are parts such as long thin fibres, that could move due to air movement during the pose.

Lastly, since we had to work with onsite ambient illumination, whose colour temperature was between 3000 K° and 3500 K°, the light sources used in the standard sets could not be used since their colour temperature is 5500 K°. In fact, in a set that includes light sources with different colour temperatures, the predominance of one kind varies depending on the surface's orientation, leading to varying and unpredictable variations of white balance. This would not affect the 3D reconstruction, but the textures, since they would incorporate the local predominant colour causing inexistent variations in the apparent colour of the surface itself, making it difficult, when not impossible, to normalise these textures and get their proper local colour. So, in such bad cases, these are the guidelines that have been set and followed: white balancing has been done in relation to the onsite illumination; objects have been kept still, and the operator moved as freely as possible around the object; pictures have been taken using long exposures due to the overall dim lighting. Instead of using shorter exposure times with high ISO to minimise the image, noise ISO has been kept in the optimal range for the cameras (100-400 ISO), and exposure times have been prolonged.

### 3.4. Digital Libraries of 3D Cultural Heritage Objects and Virtual Museums

In the past decade, there has been rapid development in the fields of 3D and Multimedia Digital Libraries and Virtual Museums[2] and recently, we have been assisting in attempts to connect Digital Libraries with 3D interactive experiences.

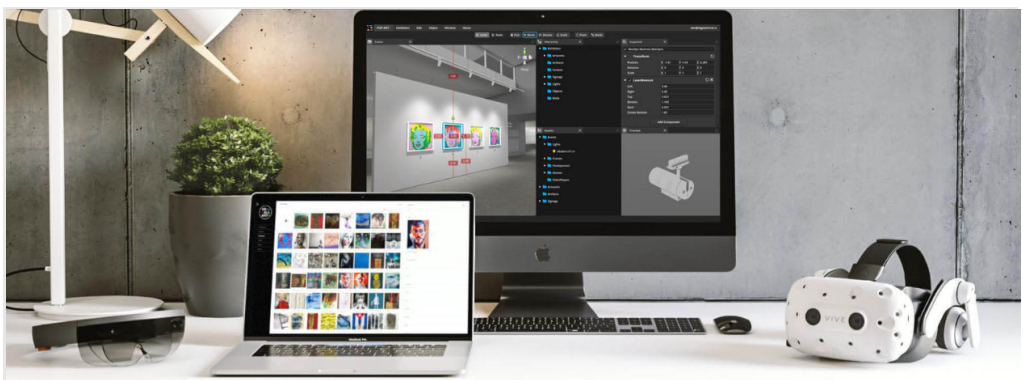

Fig. 2. VRallART approach, with the dashboard and the 3d immersive exhibition creator application - source: https://vrallart.com/.

It is the case of Occupy White Walls (OWW: https://www.oww.io/), a 3D interactive platform dedicated to artists and curators who aim to exhibit their work, inviting people into a virtual social cyberspace with their avatars for a gaming experience simultaneously. OWW is connected with Kultura (https://kultura.art/), a social media dedicated to art and powered with AI (Daisy), where users can upload their artworks digitally, together with relevant metadata, contributing to creating a large digital library that can be inquired by OWW, when creating the exhibit. The framework is mainly, but not exclusively, optimised for 2D graphics, and it is now managed by Kulturaexmachina (https://kulturaexmachina.com/).

---

[2] We focus here on Virtual Museums, intended as 3D interactive and narrative experiences (Pescarin, 2014), and on 3D Digital Libraries, meant as structured and semantically enriched collections of 3D and multimedia objects.

Similarly, but with a different approach to social participation and collaborative 3D exploration, is the VRallART framework (https://docs.vrallart.com/). A web front-end is used as a dashboard to upload and curate the artworks that can be used and included in Virtual Exhibitions specifically created with a "Virtual Editor" to enable visitors to immersively explore the artworks (Janković et al., 2020) (Fig. 2). A similar approach is used by some applications developed with the framework Open Source ATON (Fanini et al., 2021) developed by CNR ISPC.

In the Italian National framework, a remarkable example is the case of the E-Archeo project, a large-scale national project focused on 8 Italian archaeological sites, sponsored by the Ministry of Culture (MiC) and coordinated by Ales SpA (https://e-archeo.it/en/)(Pietroni et al., 2023). The project produced several online and on-site outputs for general and specialist audiences. One is e-Archeo 3D, an interactive web3D application based on the framework Open Source ATON developed by CNR ISPC (see *infra*). The App (https://3d.e-archeo.it/a/ales/) allows the visualisation of the digital replicas and virtual reconstructions of 8 archaeological sites from the north to the south of Italy through 360-degree interrogable scenarios provided with multimedia contents like videos, images, 3D models of individual artefacts (Fig. 3). Furthermore all the digital contents have been published on the Zenodo communities and are available for download (https://zenodo.org/communities/?p=e-archeo). This example highlights the importance and the opportunities offered by the digital transition process and the value of 3D models in Cultural Heritage dissemination.

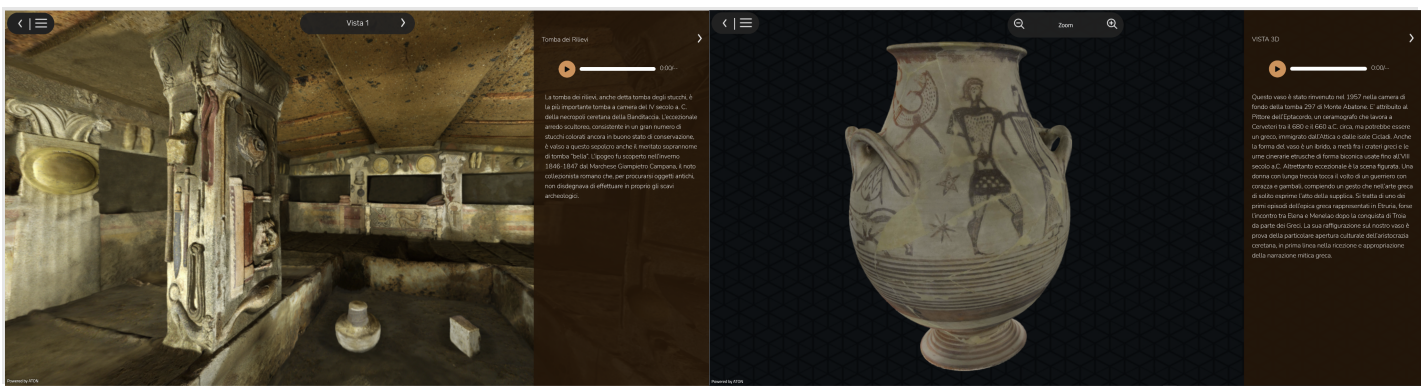

Fig. 3. Examples of interrogable digitised 3D archaeological sites and artefacts provided with textual and narrative information and available on the e-Archeo 3D platform (credits: e-Archeo 3D).

## 4. Methodology

In the creation of the Digital Twin of the exhibition, as a first necessary step, we planned to acquire the entire collection on display, together with the multimedia content available to the visitors and the physical spaces (the rooms).

The acquisition of most objects and environments lasted three months (from the end of March 2023 to the end of June 2023). For the entire duration of the acquisition, we dealt with specific time constraints, mainly related to the fact that we had to work on an ongoing temporary exhibition that closed at the end of May. Once the exhibit was over, the displays would be dismantled to return to the original structure of the visitor's route. In addition, not all objects belonged to the permanent collection of Palazzo Poggi but were on loan from different museum/cultural realities, as often happens to temporary and thematic exhibits. For this reason, the acquisition timeline was dictated by the fact that, even before the end of the exhibition, some objects had to be returned to the owner institution they belonged to.

Having to work while the exhibition was still up meant that access to the premises was restricted. The only day of the week available for the acquisition campaign was Monday, the day of Museum closure. In particular, we worked from 9:00 to 16:00 for nine days. In addition, we had the chance to access the museum at the same time on four other days, thanks to the fact that one of the rooms was closed for three weeks to enable the curators to reorganise the spaces for the new exhibition. We spent a total of 91 hours for the entire acquisition process.

The work was carried out by 6 acquisition teams, the staff of 3 museums and one library and specialised external professionals, specifically:

- researchers and students from five Departments of the University of Bologna (Architecture; Civil, Chemical, Environmental, and Materials Engineering; Classical Philology and Italian Studies; Cultural Heritage; History and Culture);
- researchers from the Digital Heritage Innovation Lab of CNR ISPC (Florence and Rome branches);
- managers and staff of the University of Bologna Museum Network;
- managers and staff of the Bologna University Library;
- staff of the Archaeological Museum of Bologna;
- staff of the Medieval Civic Museum of Bologna;
- professionals specialised in the removal and reinstalling of display cases.

To reach the project objectives, we have identified and set up a common protocol that would have guided us. Before starting this acquisition, to avoid wasting working effort, we assessed all the pieces included in the exhibition. We checked whether they had already been acquired in the past years in the context of particular research projects within the University of Bologna. A few of them were in this situation and, once they got permission to reuse them, were removed from the acquisition schedule.

We organised the acquisition of the remaining pieces focusing first on the objects that were borrowed from other cultural heritage institutions and that had to be returned. For some of these objects, it was compulsory to acquire them together with a representative staff of the source institution, who should assist the process. For a few of them, we agreed on postponing the acquisition once they were returned to the source institution, running the acquisition process there.

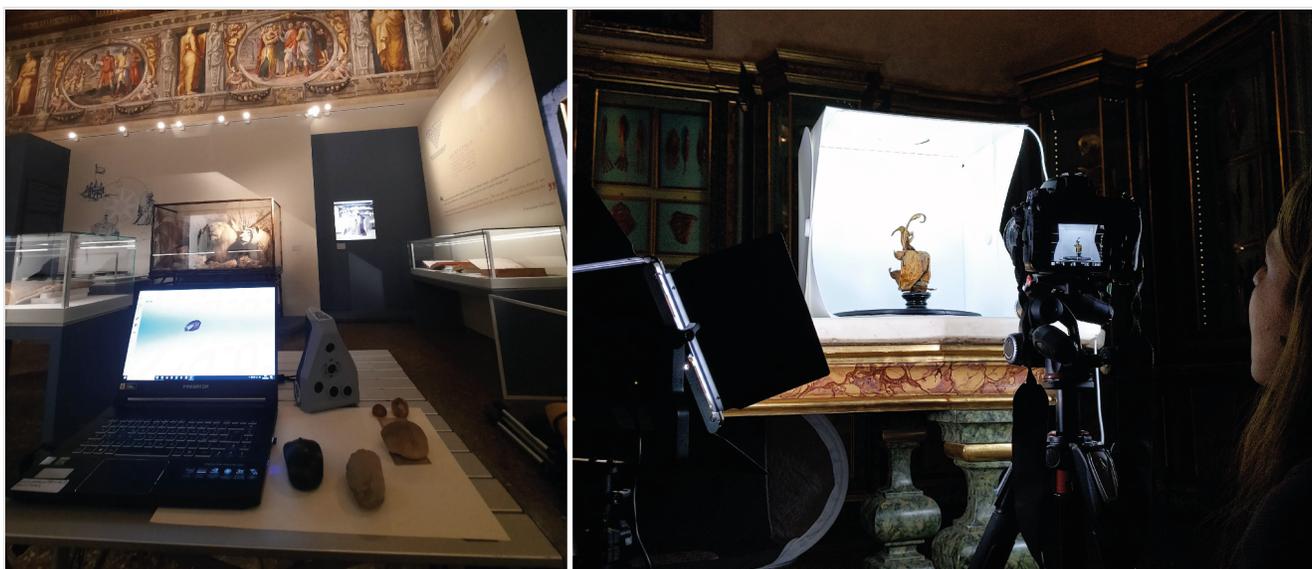
Fig 4. Researchers acquiring different types of objects using a structured light projection scanner (left) and photogrammetric techniques (right) and using supports such as a camping table (left) and a marble table (right).

Then the acquisition proceeded with the objects positioned inside weighty display cases that required specialised professionals to be removed and reinstalled – an expertise we had available only on a few dates. For three of these objects that depicted non-unique pieces (such as common animals) and were not easy to acquire considering such constraints, we decided to re-use already existing 3D models available, with appropriate licences, on specific platforms (e.g. SketchFab) and use them as representative objects of those shown originally in the exhibition. Then, we moved to acquire all the other objects of the exhibition that are usually preserved by the University of Bologna Museum Network and the Bologna University Library. Due to the time constraints mentioned above and the need to move the objects from the display cases they were positioned to perform the task, the acquisition required strong coordination among all groups working on the project.

A further constraint encountered by the research groups was the physical space. For the same reasons already described in Section 3, it was challenging to organise the acquisition activities of several teams in a location such as Palazzo Poggi. This was mainly due to the large number of showcases in the rooms, the numerous technical equipment/instruments brought by the groups, and the scarcity of electrical sockets and supports. On this last point, as shown in Fig. 4, we have used camping tables as temporary stations for the acquisition tasks, which enabled us to work in the various rooms used for the exhibition, thus minimising the distance from the display case where an object was positioned to the station where we relocated it to perform the acquisition. In addition, on Mondays, we also had available two marble tables in other rooms of the Museum, which allowed us to work on an ideal setting for acquiring some of the objects.

Each team was required to collect their workflow procedures and processes and update a shared file. They did this by listing the metadata in a shared spreadsheet on a cloud-based service (Sharepoint). This methodology was chosen for two main reasons. First, it allowed for the division of labour, with each group concurrently gathering their specific section of the required workflow metadata. Second, storing the spreadsheet in the cloud allowed multiple contributors to make real-time updates and contributions simultaneously. This resulted in an efficient compilation of the comprehensive metadata table, with input from each group. Additionally, the cloud platform provided features for peer review, enabling members to examine each other's work and an open environment that supported the transparent exchange of information. The metadata table was directly related to the various activities and was organised as a spreadsheet with multiple separated columns according to the different stages to be performed on each object, described in the following subsections.

The shared space was set up using Microsoft Sharepoint. This choice was convenient for the group considering that both the University of Bologna and the National Research Council adopted it to guarantee that the activities performed by their researchers are compliant with the rules introduced by the General Data Protection Regulation (Regulation (EU) 2016/679) in particular with gathering, using, and sharing personal data.

*4.1 Acquisition (step 1) and following software activities (steps 2-7)*

The initial stage of the process (step 1) aimed to capture and acquire analogue materials to create their digital representation. The information related to the acquisition phase that was tracked systematically by all the participants using the spreadsheet document included:

a) *Institution* – the organisation that is responsible for initiating and funding the digitisation project;
b) *People responsible acquisition* – a person or team of people who are responsible for carrying out the acquisition;
c) *Acquisition technique* – the specific method or approach used to capture the analogue materials and convert them into digital form;
d) *Acquisition tools* – the hardware tools used to perform the acquisition activity;
e) *Acquisition date* – the specific date or range of dates when the acquisition process was initiated or completed.

Such an acquisition phase was then followed by several *software activities* (steps 2-7), i.e. the series of phases that involve using various software tools and applications to process, transform, and publish the digital versions of the material acquired during the acquisition phase. We decided to keep and store three levels for each item: the raw data (or level 0 model), obtained directly from the acquisition software with no interpolation or holes closure; Second, the high-resolution model obtained with interpolation aimed at producing a complete mesh (level 1 model); and the optimised model for real-time online interaction (level 2 model). Although the software activities can vary depending on the nature of the materials being digitised and the intended use of the digital files, our tracked work included the following phases in the digitisation process of all objects:

- *Processing phase* (step 2) – software tools are used to process and manipulate the digital files produced during the acquisition phase;
- *Modelling phase* (step 3) – software tools are used to create a digital model of the object or space;
- *Optimization phase* (step 4) – software tools are used to optimise the digital files for specific purposes or use cases;
- *Export phase* (step 5) – software tools are used to export the digital files in a specific format or for a specific purpose;
- *Metadata creation phase* (step 6) – software tools are used to create structured metadata that describes the content and context of the digital files
- *Upload phase* (step 7) – software tools are used to transfer digital 3D models from a local device or storage location to a Web-based framework (e.g. ATON).

Similarly to the acquisition, each software activity is characterised by various elements that are important for tracking and documenting the digitisation process:

f) *Institution responsible for the activity*;
g) *People who took part in the activity;*
h) *Tool employed for the activity;*
i) *Dates in which the activity took place.*

These metadata are necessary to reconstruct and describe, for each object, a detailed record of the entire digitisation process, allowing researchers to track the progress of the digitisation process. Moreover, by documenting these elements for each software activity, institutions can maintain a comprehensive record of the digitisation process, which can be useful for tracking progress, evaluating the project's success, and ensuring the long-term preservation and accessibility of digital materials. In addition, it would also allow further study on the dynamics of the acquisition and digitisation process, which could take from a few hours to several weeks, depending on the specific characteristics an object has.

*4.2 The objects in the temporary exhibition*

The objects in the exhibition "The Other Renaissance", totalling 301, belong mostly to the permanent collection of Palazzo Poggi and, to a small extent, have been loaned by cultural institutions such as the University Library of Bologna (BUB), the Library of the Department of Biological, Geological and Environmental Sciences - Bertoloni Historical Fund, the Library of Mathematics, Physics and Computer Science - Physics Section - "Guido Horn D'Arturo" Historical Library, the Department of Biomedical and Neuromotor Sciences, the Medieval City Museum, the Archaeological City Museum and the State Archives of the city of Bologna, the Carrara Academy of Bergamo, the Museum of Civilisations and the Spada Gallery of the city of Rome, the Academy of Physiocritics of Siena, and finally the Natural History Museum of Verona. The assets were distributed in six rooms and presented a wide variety (as shown in Tab.2).

Objects displayed very diverse characteristics, both in their geometric shapes and surface properties, adding a further constraint to those already mentioned above (i.e. time and space). Digitisation involved many materials, including paper-based objects such as manuscripts, printed volumes, and ancient maps. Moreover, there were numerous woodcuts, technical/scientific instruments, statues, specimens, and archaeological finds. To ensure their preservation, animal models were treated with substances like shellac, often incorporating fibrous materials to repair damage (Reggiani, 2022). These treatments increased the complexity of the materials, resulting in many cases of irregular and reflective surfaces. Notably, Room 5 exhibited an extensive and varied collection of natural artefacts, varying in size and materials, such as large marine turtle shells, different amphibian and cartilaginous fish specimens, as well as numerous other geo-paleontological artefacts like rocks samples and minerals, fossils, and microfossils. Handling such a vast and diverse collection required the application of established techniques. As a result, the acquisition process relied on well-tested remote-sensing technologies commonly used in cultural heritage preservation (and thus available to the various research groups), such as 3D structured light projection scanning and digital photogrammetry (Bitelli et al., 2007; Apollonio et al., 2021). The most appropriate acquisition technique was influenced by the properties and constraints mentioned above, in particular, surface type, size, collocation, and geometric complexity (Peinado-Santana et al., 2021). The use of image-based techniques, like digital photogrammetry, offered certain advantages. Indeed, structured light projection scanners have strict requirements regarding acquisition distance range. For this reason, it was necessary to move around the object while keeping the correct instrumental distance, which was not always possible when in the presence of obstacles or limited spaces. These ideal acquisition conditions were difficult to reproduce when the objects could not be moved from their original locations. In some cases, it was necessary to adopt a photogrammetric approach (Ruiz et al., 2022). This image-based technique was more adaptable in the acquisition distance than the structured light projection technique. Moreover, using a close-range approach to photogrammetry, images were recorded at a considerably higher resolution, thus resulting in better reconstructed 3D models. Additionally, photogrammetry was effectively used to capture irregularly shaped objects (Remondino, 2011). In the case of the Aldrovandi collection, there were, for example, many animal models characterised by jagged edges and very accurate details. Moreover, the implementation of photogrammetry helped to lower acquisition costs and time, as it required easily accessible tools like digital cameras (Guidi et al., 2015). For their part, structured light projection scanners, despite the high cost of the equipment, allowed for fast and accurate retrieval of a large amount of data in a very short time. Surfaces were measured and processed by providing 3D coordinates of points in real time (Modabber et al., 2016). Furthermore, depending on the type of object to be acquired, the teams could choose among different scanner models suitable for different sample sizes. That made it possible to quickly collect a large number of assets. Another advantage these range-based systems offered was their handheld configuration, which made the operational phase of scan acquisition more practical (Georgopoulos et al., 2010). The careful choice of which technology to use on the different objects in the Aldrovandi collection allowed the research groups to proceed simultaneously and quickly with the digitisation process. Moreover, the limitations we had to deal with (time, space, and material) proved valuable as they allowed us to test the effectiveness and efficiency of widely used acquisition technologies. They also helped to address the logistical challenges that commonly arise during the intricate digitisation process.

As listed in Table 2, we have acquired 104 specimens, 27 printed volumes, 17 manuscripts, 5 nautical charts and maps, 1 diorama, 7 herbariums, 21 models, 7 woodcuts, 3 paintings, 6 painted ceilings, 11 casts, 2 medals, 4 scientific instruments and more than 30 other objects, including archaeological remains; among multimedia assets, we have acquired 9 videos, 2 prints and 27 panels with graphics.

| Room n. (n. of objects) | Types of objects (n. of objects) |
|---|---|
| 1 (30) | Video (2), Specimen (8), Printed volume (6), Nautical chart (3), Print (1), Diorama (1), Herbarium (1), Manuscript (1), Painting (1), Rooms/Painted ceilings (1), Panels with graphics (5) |
| 2 (39) | Herbarium (6), Printed volume (4), Manuscript table (4), Specimen (4), Model (3), Manuscript volume (2), Map (2), Paintings (1), Vase (1), Rooms/Painted ceilings (1), Video (1), Panels with graphics (10) |
| 3 (20) | Woodcut (7), Printed volume (5), Video (2), Manuscript table (1), Rooms/Painted ceilings (1), Panels with graphics (4) |
| 4 (13) | Video (2), Knife handle (2), Printed volume (2), Mask (1), Pendant (1), Rooms/Painted ceilings (1), Panels with graphics (4) |
| 5 (146) | Specimen (79), Artifact (23), Cast (9), Gemstone (9), Manuscript volume (8), Printed volume (5), Medal (2), Statue (2), Video (2), Manuscript table (1), Necklace (1), Rattle (1), Lamp (1), Axe (1), Print (1), Rooms/Painted ceilings (1) |
| 6 (53) | Model (18), Specimen (13), Printed volume (5), Cast (2), Print (1), Illuminated manuscript (1), Manuscript table (1), Painting (1), Microscope (1), Compass (1), Bottle (1), Electrostatic machine (1), Discharge arc (1), Technical instrument (1), Rooms/Painted ceilings (1), Panels with graphics (4) |

Tab. 2. The rooms, types and number of objects in the "The Other Renaissance" exhibition.

## 4.3 Tracking the exhibition's original narrative via Virtual Tours

During the acquisition campaign, we soon realised that, in a very short time, the entire exhibition would not exist anymore, while during data post-processing, we would have needed to check objects' position and relation with the guided narrative.

Moreover, some of the items were given back to owner institutions before the closing of the event. We therefore decided to design and develop a simple mock-up of the exhibition to be used as a common and shared reference during the creation of the Digital Twin. Due to the mentioned time constraints, we have chosen as a viable solution 360 panoramas Virtual Tour (VT) technology.

Today, although there are different types of Virtual Tours and approaches (Pescarin et al., 2023b), 360 VTs are widely adopted in different domains and for different purposes, such as educational, tourism (real estate agency sector), and business growth by different types of companies. They are common tools to immediately and digitally connect with physical places, even remotely, and can provide several types of information thanks to their capability of being enriched by multimedia content (video, texts, sounds). Moreover, some of these tours are used with VR immersive headsets to increase the sense of presence and stimulate engaging experiences. In the cultural heritage domain, especially after the emergency linked to the COVID-19 pandemic (Amir et al., 2021), museums have strengthened their educational and cultural position by focusing on technologies such as VTs (Wu, 2022). In the case of the Aldrovandi exhibition, we have used 360-degree spherical or equirectangular images acquired in each room (Fig. 5), making this the most suitable solution to provide complete information.

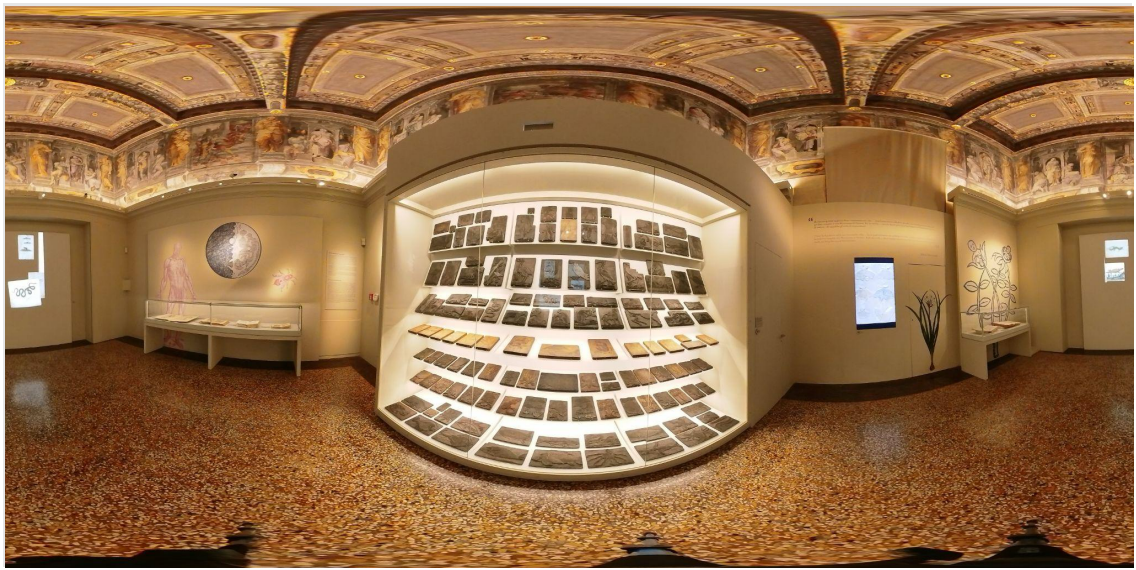

Fig. 5. Examples of spherical or equirectangular images from the Room 3 of the exhibition.

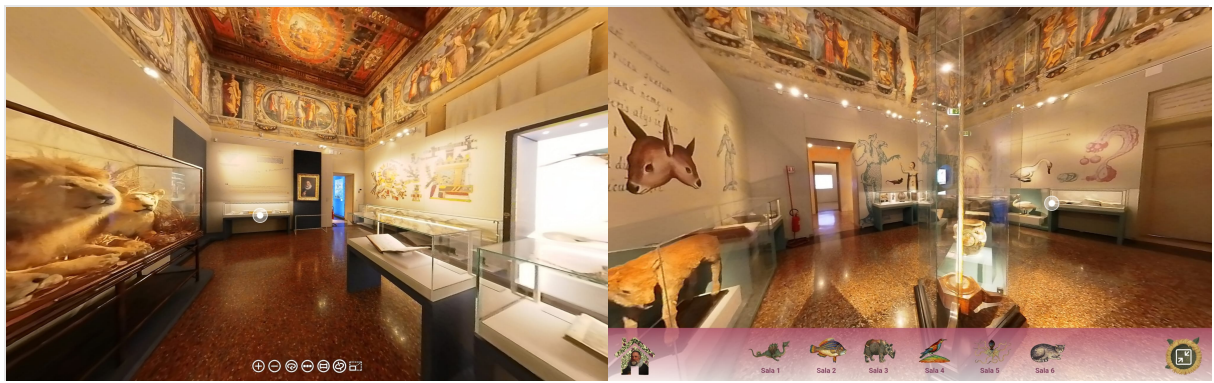

Fig. 6. Two different UI of the virtual tour navigation mode, designed by the Skin Editor tool.

Since we did not need to obtain super-high-resolution images, we adopted a fast technique, employing an Insta360 ONE X2, a dual-lens 360 action camera conceived for capturing panoramic videos and images up to 5.7 K. We planned to acquire more than one panorama in each room, to be able to virtually visit the exhibition following the usual visitor path and the audio guide that was available for the visitors during the exhibition. The height of the camera was set to simulate the visitor's eye. During the acquisition, we could immediately check the result on a tablet from the Insta360 Studio mobile app connected through Wi-Fi to the camera, thus minimising the time spent and avoiding potential mistakes. The images have been edited and exported to the maximum resolution available, obtaining 360-degree images of 6080 x 3040 pixels. All single panoramas have been uploaded to the common shared cloud.

The different shots have been interconnected to simulate a guided and one-directional navigation of the spaces. This work, creating the proper interactive Virtual Tour, was carried out using Pano2VR (https://ggnome.com/pano2vr/). Its main function (Cao, 2022) is to convert panoramic images into formats suitable for web-based views (HTML5/CSS3) on different platforms. By importing the panoramic images into the software, it is possible to link them in a path that reproduces the exhibition path. The tour can be enriched with interactive elements such as informative pop-ups, photographic hotspots, directional audio, and video. The UI is customisable through simple CSS to define the formatting and visual style, thanks to the Skin Editor tool (Fig. 6).

## 5. Acquisition and Processing

In this section, we describe our direct experience while acquiring and processing the objects of Aldrovandi's exhibition. In particular, we detail the work done with structured light scanners (Section 5.1) and photogrammetry techniques (Section 5.2). Then, we discuss some lessons learnt and challenges addressed during the process (Section 5.3), and, finally, we introduce the approach followed for recreating the rooms where the exhibition was originally set (Section 5.4).

### 5.1 Structured light projection scanning

Structured light projection scanning (SLS), a high-precision survey methodology based on the projection of a light pattern of non-coherent and diffuse radiation onto the objects' surface, was employed for the 3D survey of a consistent number of objects in the exhibition. SLS technology was considered a suitable geomatic technique for the aims of this project because of its ease of use, its speed of acquisition and its accuracy in detecting very detailed geometric features such as those presented by numerous objects in the Aldrovandi collection (Bitelli et al., 2020; Forte et al., 2023). The Structured light projection 3D survey was performed employing two handheld scanners from Artec company: the Artec Space Spider scanner and the Artec Eva, both associated with the software Artec Studio Professional and useful for the acquisition of objects belonging to two different size range (Franco de Sá Gomes et al., 2019; Moyano et al., 2023). The technical specifications of the two scanners can be found in Tab. 3.

| Scanner | 3D point precision | 3D resolution | Texture resolution | Acquisition surface | Acquisition distance |
|---|---|---|---|---|---|
| Artec Space Spider | 0.05 mm | 0.1 mm | 1.3 MP | 90 x 70 mm<br>180x140 mm | 20-30 cm |
| Artec Eva | 0.1 mm | 0.2 mm | 1.3 MP | 214 x 148 mm<br>536 x 371 mm | 40-100 cm |

Tab. 3. Technical specifications of two Artec structured light projection scanners.

To conduct the SLS survey for this work, two Spider scanners and two Eva scanners owned by the Department of Cultural Heritage (DBC) and the Department of Civil, Chemical, Environmental, and Materials Engineering (DICAM) of the University of Bologna were used. For both scanners and groups, the methodology employed was the same. Regardless of whether the objects could be moved or not (some were too heavy, some too small or too fragile, so the decision was to acquire only the parts outwards and not those facing the support or the showcase), they were acquired by scanning until the complete obtainment of the geometric information regarding the accessible surface (the mode is shown in Fig. 7), being careful to maintain an overlapping area between the scans (Braren and Fels, 2020; Ugolotti et al., 2022).

In the typical workflow (adopted for most cases with some exceptions for a few peculiar objects – see section 5.3), each scan obtained was processed using the proprietary software Artec Studio Professional (the same used for the scan acquisition). The first step was to check for the quality of the data acquired, then the single rough scans have been "cleaned", registered together, and subsequently aligned reciprocally selecting homologous points. Finally, after applying a noise removal algorithm to clean the data, the scans were merged to obtain a solid, consistent 3D mesh (level 0). The final steps consisted of closing the holes (if present) in the solid surface and applying a mesh simplification to reduce the number of polygons (level 1), necessary step given the presence of extremely high-poly meshes. We opted for a stronger decimation for larger objects with respect to smaller ones (in terms of object sizes). The range of polygon numbers was from 500,000 to 1,000,000 circa. As an ultimate step, textures were applied with a maximum resolution of 16,384x16,384 – the dimensions of the raster file of the texture map. The tool provided by Artec Studio was used for minor texture retouches, while more complicated editing was conducted with Adobe Photoshop. In any case, the export format of the 3D models was OBJ (coupled with the Material Texture Library MTL file and the Texture map PNG) and never exceeded 800 MB per 3D model.

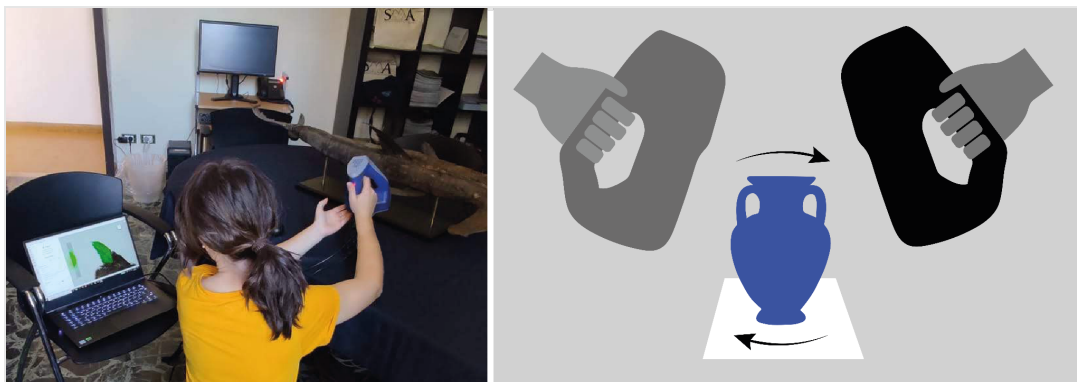

Fig. 7. Researcher acquiring with an SLS technology and the acquisition modality adopted.

## 5.2 Photogrammetry techniques

Using photogrammetry techniques, as for structured light scanner techniques, allows a complete and flexible description of an object's morphology without direct contact (Bitelli, 2002; De Luca, 2011). Photogrammetry is a well-known and common technique today (Kraus, 2007; Guidi et al., 2010; Russo et al., 2011). Regarding the objects of the Aldrovandi exhibition acquired with photogrammetric techniques, as summarised in Table 4, five cameras were used for the acquisition task, owned by the Department of Classical Philology and Italian Studies (FICLIT), the Institute of Heritage Science at CNR (CNR ISPC), the Department of Cultural Heritage (DBC) and the Department of Architecture (DA).

Different instrumentation and configurations were used to obtain morphologically accurate models with high-detail textures. These objects vary not only in formal and material characteristics but also in their placement. In particular, the acquisition setup had to be adjusted based on the mobility of the exhibited artefacts. The acquisition schemas were employed as follows (and summarised in Fig. 8) to allow faster processes and uniform and well-diffused illumination to be performed (Webb et al., 2020):

1. for small-size movable objects, we used lightbox, turntable, fixed camera on a tripod;

2. for medium size movable objects, we used 2-4 continuous led lights on stands, photographic studio background, turntable and fixed camera on a tripod;

3. for big size or non-movable objects, we used 2-4 continuous led lights on stands and cameras on tripods (moved around the object).

| Group | Cameras | Sensor size | Image resolution | Pixel size | Focal length |
|---|---|---|---|---|---|
| FICLIT | Panasonic DMC-LX100 | CMOS 17.3x13mm | 4112x3088px | 4.19 μm | 24-75 mm |
|  | Nikon D7200 | CMOS 23.5x15.6mm | 6000x4000px | 3.89 μm | 50 mm |
| CNR ISPC | Canon EOS 6D | CMOS 36x24mm | 5472x3648px | 6.54 μm | 50 mm |
| DBC | Nikon D750 | CMOS 36x24mm | 6016x4016px | 5.95 μm | 40-70 mm |
| DA | Sony A7 I | CMOS 36x24mm | 6000x4000px | 5.93 μm | 28-70 mm |

Tab. 4. Technical specifications of the cameras used by the acquisition groups.

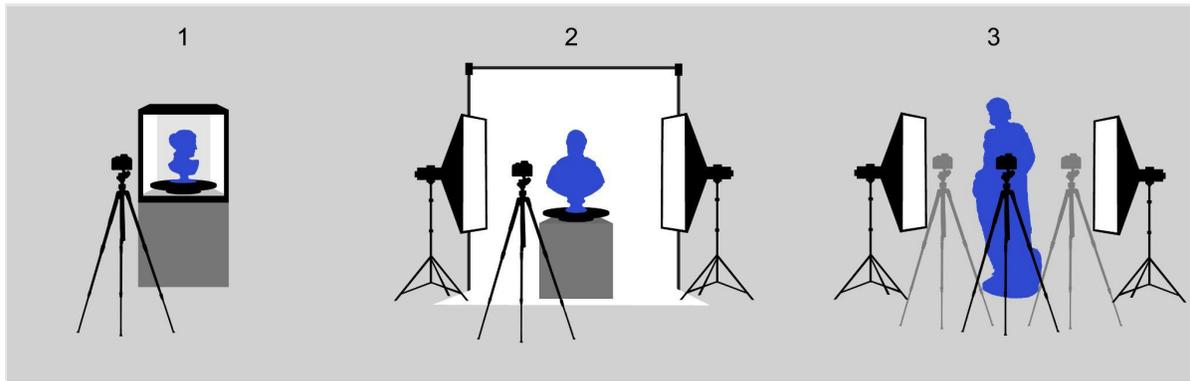

Fig. 8. Acquisition schema and shooting techniques adopted for surveying the museum collection.

During the acquisition, the objects were positioned at the centre of a platform with a coded target grid to improve camera orientation and model scaling (Guidi et al., 2010; Sapirstein, 2018; Luhmann et al., 2020). The camera was placed frontally at an appropriate distance to ensure sufficient depth of field and focus on both the nearest and farthest points of the objects. For objects with significant depth variations, a narrow aperture and a fairly large focal length were selected for a greater depth of field. During the acquisition phase, the platform was rotated at angular intervals of 30 degrees to ensure overlap between adjacent shots (Lo Brutto and Spera, 2011; Menna et al., 2017). This close-up image configuration simulated a shooting geometry with many intersections, resulting in a dense and geometrically reliable network (Luhmann et al., 2020). The camera was also vertically translated during each rotation to capture data necessary for reconstructing each part of the object (Collins et al., 2019). To achieve accurate colour data and exposure settings, an X-Rite ColorChecker Passport Photo target with 24 patches of known reflectance values was used (Apollonio et al., 2021).

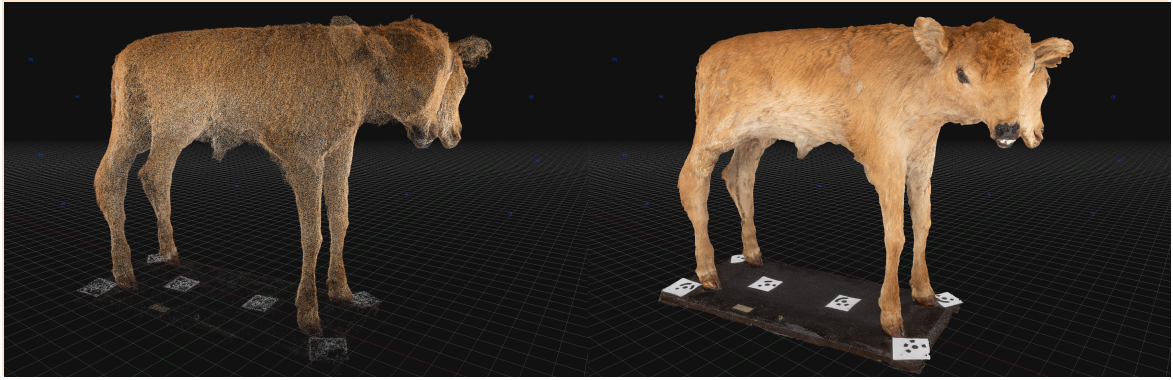

Fig. 9. Example of a 3D model after photogrammetric computation of the dataset: left) dense point cloud; right) textured mesh.

The subsequent data processing was performed in Structure from Motion (SfM) software and followed the typical pipeline (exemplified in Fig. 9), including the extraction of homologous points, camera orientation, object scaling, dense point cloud creation, mesh generation, and projection of photographic textures onto the model (De Paolis et al., 2020). Agisoft Metashape and 3DF Zephyr have been used, while a testing phase on the open-source Meshroom based on the AliceVision framework is in progress (https://alicevision.org/#meshroom).

*5.3 Lessons learnt from the object acquisition and processing under certain constraints*

This section introduces the main risks we encountered while acquiring and processing the objects in the exhibition. We also describe the solutions we adopted for minimising such risks, as summarised in Table 5.

5.3.1 Acquisition of non-Lambertian materials

Acquisition based on photogrammetric techniques is best suited for opaque non-reflective matte surfaces or Lambertian materials, for which it is assumed that every point of their surface reflects incident light in all directions with the same intensity, thus making its illuminance independent of the viewpoint of the observer. Non-Lambertian opaque materials present are more complex to be treated. Structure from Motion (SfM) algorithms, in fact, first have to recognise the same point of a surface in different images of the dataset to build the 3D model and then struggle if its colour varies depending on light or view incidence angle, as it happens with non-Lambertian opaque materials, because of their reflective surfaces. Perfectly Lambertian surfaces are an abstraction, but the reflective component and the Lambertian can be treated separately, and hence, if we could isolate the specular component and eliminate it, we could then effectively use the images within the standard photogrammetric workflows. Cross Polarisation is the standard solution to accomplish this task (Frost et al., 2023). This technique uses light sources screened with polarised filters and another polarised filter applied on camera lenses to eliminate most specular reflections and leave the diffusive component almost intact when the image is taken. But for this technique to work properly, some requirements are needed: (a) there should be no other light sources than the polarised ones hitting the object, and (b) it only works with direct light, so the environment should be as dark as possible.

| Nr | Risk | Solution(s) |
|---|---|---|
| 1 | Acquisition of non-Lambertian materials | • Cross Polarisation – creating a set-up with light sources screened with polarised filters.<br>• Minimise effect on the reconstruction process by lowering the illuminance of light sources using white lightboxes, substituting the whole environment with his hard light sources with a self-illuminated white box where the light intensity is almost evenly spread on all the surfaces. |
| 2 | Impossibility to move some | • Interposing white fabric screens between light sources and object, with all ambient lights shut |
| | objects and to open glass cases | down and light sources positioned on opposite sides of showcase, high above its top surface, with an inclination of 45° (emitted light hit the fabric sheets and generated a lightbox effect) and black fabric curtains placed behind the cameras to avoid reflections on the glass (cameras remotely controlled). |
| 3 | Materials with no intrinsic proper colour (clear transparent or translucent materials and polished metals) | ● Glass: techniques used by Karami et al. (2022).<br>● Translucent: cross-polarisation techniques (Angheluta and Radvan, 2020).<br>● Computer Graphics modelling – if geometry is simple enough, measuring with simple tools like a metre and calliper and then direct modelling in 3D. |
| 4 | Presence of physical colours | ● Cross Polarisation and avoidance of too different acquisition positions. |
| 5 | Complexity in size and details (i.e. hairs, fur, fibres, frayed edges, thorns, spikes, and teeth) | ● Computer Graphics modelling – direct intervention on the model to optimise details.<br>● For sharp objects with a small diameter but a considerable height, segment the data by slicing portions horizontally. |
| 6 | Time constraints versus object complexity | ● Luhmann et al. (2020) formula.<br>● High overlap of acquired photos.<br>● Lightbox set with intermediate intensity (in case of reflections).<br>● Microstructural details acquired setting larger aperture (Nicolae et al., 2014). |

Tab. 5. The risks and related solutions adopted in the acquisition and processing of the exhibition objects.

In our acquisitions, these limitations would have required thoughtful planning of the set and its isolation to eliminate all ambiental interactions, an operation which, given the operational conditions, was deemed impractical due to the logistic and time limitations. Since there were no viable solutions to eliminate specular reflections, the strategy has been to minimise their effect on the reconstruction process by lowering the illuminance of light sources using white lightboxes. A lightbox allows substituting the whole environment with his hard light sources with a self-illuminated white box, where the light intensity is almost evenly spread on all the surfaces apart from the opened one in the direction of the camera and the bottom one unless the lightbox is placed on a transparent surface and lit from below as well. Specific attention had to be paid to the placement of the object inside the lightbox because one of the sides is opened to allow the object to be shooted, and, through this window, there could be a direct line between a hard light of the room and the object. So, the object should be placed well inside the lightbox, and any hard light source that could not be avoided had to be cloaked.

An example of such a problem has been acquiring a pre-Columbian mask. In this case, we had to repeat the acquisition due to a screen placed above it in the exhibition showing an animation with information and details about the object. Even if, by the naked eye, its contribution could not be spotted in the images, the dim but variable light source introduced in the first dataset had fine variations, which obliterated a smooth reconstruction process. By using this setup, the hardness of specular reflections is greatly reduced, as well as the risk that an image would capture overexposed highlights. The drawback is that specular reflections become almost indistinguishable by proper colours, and textures will lose contrast and be lighter, but the 3D reconstruction is greatly eased and made more precise.

5.3.2 Impossibility to move some objects and to open glass cases

When the object could not be moved, the same effect could be obtained by interposing fabric screens between light sources and the object. This was the case of the three taxidermied lions, shown in Fig. 10. This group could not be removed from their original historical showcase, so we had to use a structure on whose sides white fabric sheets were placed. All the ambient lights were shut down, and two illuminators were positioned on the opposite sides of the showcase, high above its top surface with an inclination of 45°. Emitted light hit the fabric sheets and generated the effect of the lightbox described earlier.

This specific case posed another problem since the display case was made of glass, which is a reflective material. Any external light source, direct or indirect, would generate mirror reflections on the surface of the glass panels posed between the camera and the object, and these reflections would have compromised the 3D reconstruction. The problem was solved by eliminating anything that could generate these reflections by placing a black fabric curtain behind the camera and cloaking it and its tripod with another black fabric sheet. The camera was then operated remotely with a controller.

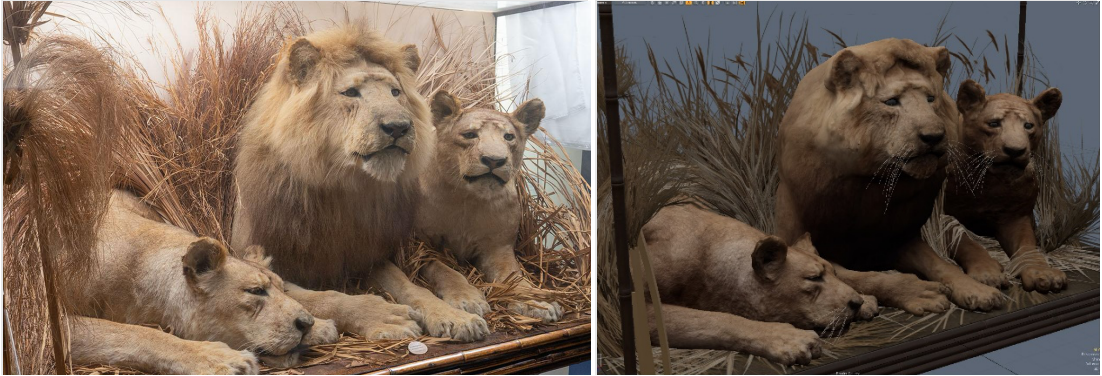
Fig. 10. Three taxidermied lions and the final textured 3D model.

### 5.3.3 Materials with no intrinsic proper colour (clear transparent or translucent materials and polished metals)

Problems with non-Lambertian materials could be solved either by choosing SLS over SfM technology or, when choosing SfM photo-based technology is a constraint, adopting ploys to reduce hard specular reflections and by doing so getting a better reconstructed 3D model. However, other materials that were part of some objects could not be acquired without using coatings. Whenever a certain point on a surface did not have an intrinsic proper colour, either SLS or photo-based techniques and instruments did not work because they did not have data to work on.

*Clear transparent materials* like glass do not reflect the incoming light diffusely, show some specular reflections and have no texture of their own to be used in image-based matching procedures. Furthermore, refraction effects can produce distortion in their appearance. Several techniques could be applied (Karami et al., 2022), but were not a viable solution for the exhibition.

In *translucid materials* like frosted glass, wax, turbid liquids, etc., most of the incident light is scattered in the thickness of the material in all directions, in either concordant (forward) or discordant direction (backward). This causes the apparent colour of the surface to change with light direction and the viewpoint because it is determined mainly by the sub-surface scattering of light, being the dominant behaviour compared to the reflectance of the surface. Some attempts have been made using Cross Polarisation (Angheluta and Radvan, 2020). But, as pointed out earlier, Cross Polarisation was not a viable solution.

*Polished metals* usually do not have a proper colour, and we see reflections of the environment. Thus, they are variable with lighting setup and point of view. Not even cross-polarisation would have helped us with such materials since there was almost no diffusive component that could be isolated.

Whenever one of these cases of non-collaborative materials occurred, the acquisition was impossible using either SLS or SfM approach. So, if geometry was simple enough, they were measured with simple tools like a metre and calliper and then modelled directly in 3D using pictures and measurements as a reference (Computer Graphics modelling). This has been the case with some historical measurement devices used for detecting electrostatic energy, as shown in Fig. 11.

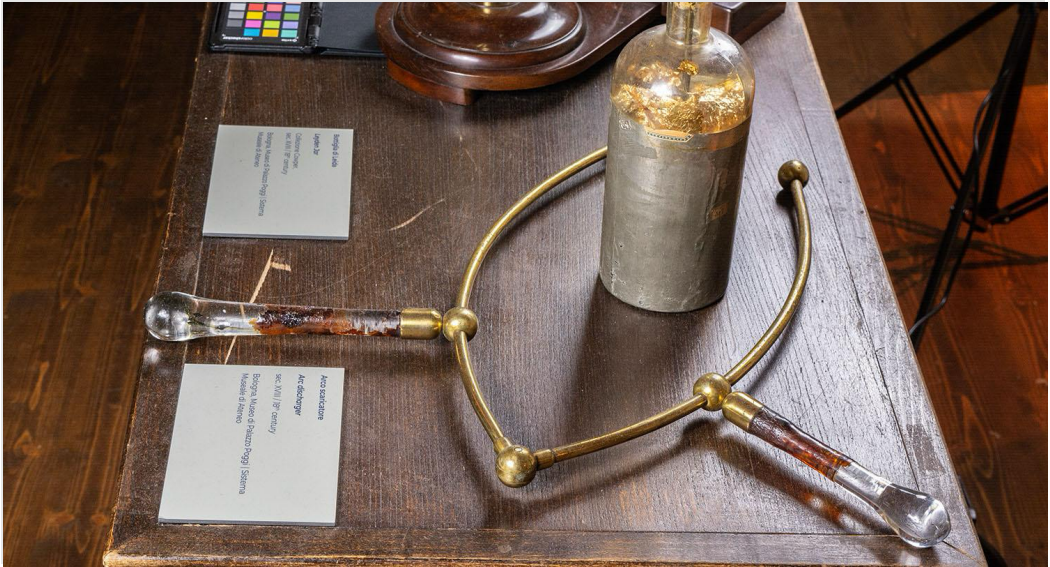
Fig. 11. Ramsden electrostatic machine (1766).

5.3.4 Presence of physical colours

A slightly different case was *physical colours,* like the iridescent metallic reflections in the feathers of some of the taxidermied birds. Physical colours are colours that change depending on the viewpoint. In nature, these typically appear on black or dark feathers and are hence the main component of the apparent colour as seen from each specific point of view, but this colour changes as the viewpoint varies, making the acquisition of such surfaces a hard task because the operator must be aware that by changing too much the view angle the colour will change causing doubled surfaces. The proper approach would have been to use Cross Polarisation but, as clarified earlier, this option was unavailable.

5.3.5 Complexity in size and details (i.e. hairs, fur, fibres, frayed edges, thorns, spikes, and teeth)

When considering materials, it is not just surface physical characteristics that pose challenges for the acquisition. There are also dimensions to be taken into account whenever the size of details is close to the resolution capability of the instrument for SLS or when it is recorded by too few pixels in the images shot for photogrammetry. Hairs, fur, fibres, frayed edges, thorns, spikes, and teeth all fit this description. There are no easy solutions for this problem since it relates to instruments' sensibility, and each case led to a different approach, even if all of them rely on a direct intervention on the model in the optimisation phase thanks to the use of 3D modelling software. Acuminated objects have been recreated, while for fabrics or fibres working with transparency maps on a simplified geometry was the most effective solution.

A noticeable example of complex acquisitions due to object size is represented by the cases of narwhal tooth ("*Unicorn*"). In this case, it was a sharp object with a small diameter (ca 5.5 cm at the base) but a considerable height (ca 225 cm). This item is stored inside a very narrow glass case (Fig. 12), which did not allow scanning inside it because of the limited space between the glass and the object, which was insufficient to ensure the correct acquisition distance for the SLS scanners (see section 5.1). Therefore, it was first necessary for specialised museum technical staff to extract the heavy object from the case and place it in an unobstructed location. In addition, the very narrow and elongated shape of the tooth and its large size in terms of height proved problematic for SLS scanning, being excessive for the instrumental range of the scanners. This required many scans (22) to ensure sufficient coverage of the entire surface of the object. In contrast to the high height of the tooth, its diameter was very small. In fact, in the uppermost parts, the tooth was isolated from the rest and could not even be placed toward a surface, both to avoid damage and because of the presence of a very wide base of support. This required the placement of objects between the tooth and the scanner to add elements within the scanned scene. Indeed, without continuity of objects within the area covered by the scanners, the real-time alignment of geometry frames is not guaranteed, and thus the scan fails. Even during processing, this case was quite complex to manage due to the object's shape and size. To perform the scan alignment, it was necessary to segment the data by slicing portions horizontally. A global alignment of an object that is beyond the instrumental range would have produced errors in the registration of the geometric frames: the frames at the base did not have common points with those at the top, and a propagation of the alignment error would have been encountered as the frames got further apart.

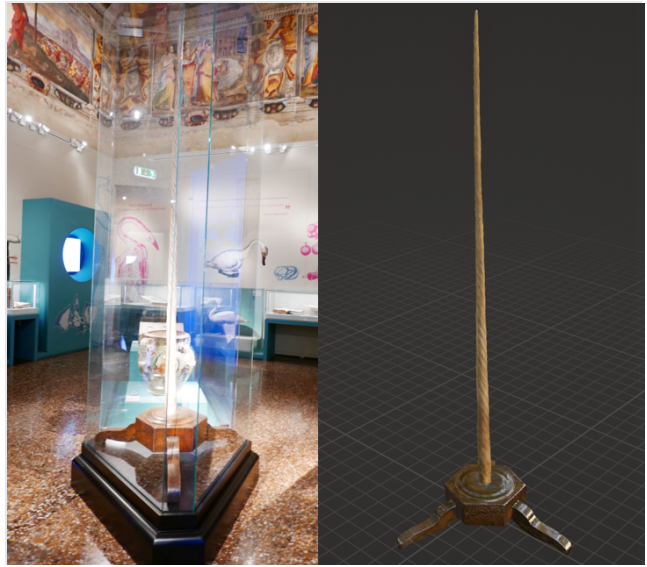

Fig. 12. The narwhal tooth in its case (left) and the final textured 3D model (right).

5.3.6 Time constraints versus object complexity

As mentioned in the sections above, acquiring some objects was characterised by other contextual constraints, such as the time to execute the entire process. In this context, it is worth mentioning the case of the Basilisk, shown in Fig. 13.

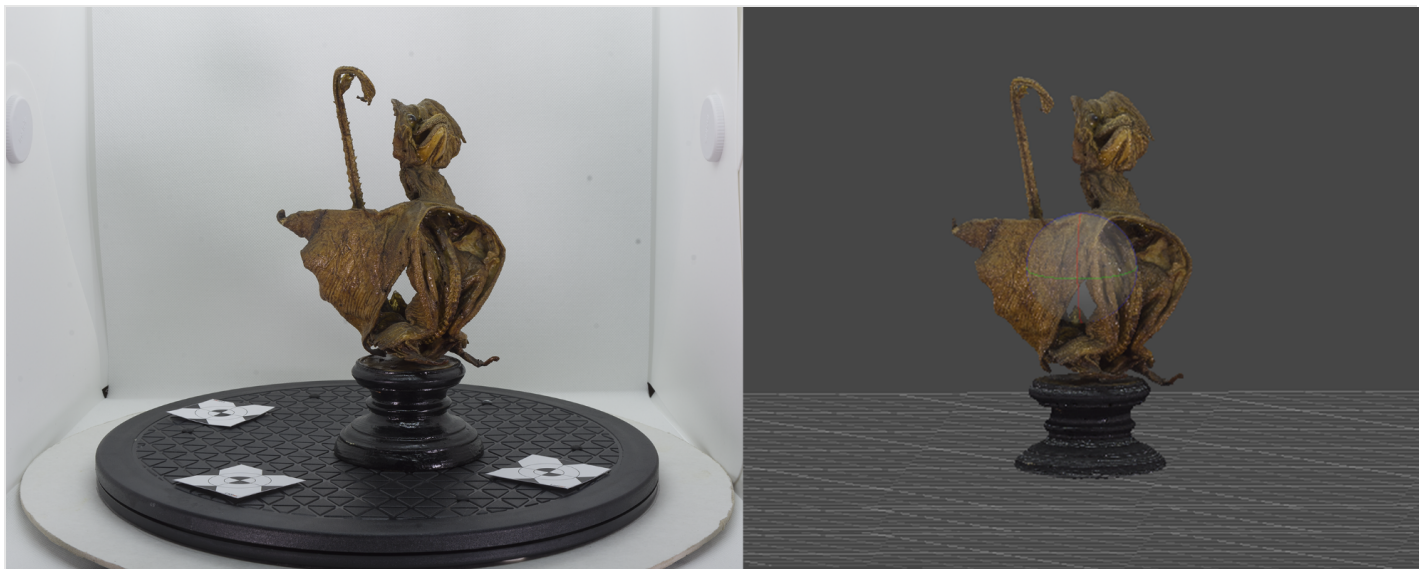

Fig. 13. Original photo and 3D model of the Basilisk.

The Basilisk, with a 21 cm height, needed to be urgently acquired as it was scheduled to return to its owner's museum within a few days from the beginning of the acquisition process. In particular, to obtain a complete model of this object, only one day was available. Such a tight time constraint would not have allowed for reacquisition in case of problems during the image alignment phase. This limitation was further compounded by the complex shape of the model. Despite having a good distribution along the principal axes and not being obstructed by external elements, it was characterised by articulated forms with subtle details and pronounced sockets that tended to create shadowed regions (Marziali and Dionisio, 2017). Certain areas, such as the junction of the tail and the lower back, as well as the coiled end of the tail and the legs, could have resulted in gaps in the 3D model due to hard shadows or areas not visible if not adequately photographed and illuminated (De Paolis et al., 2020).

To achieve a Depth of Field (DoF) that allowed focusing on the turntable 30 cm in diameter, the camera, a Nikon D750 with a Nikon AF 28-105 f/3.5-4.5 lens, was positioned at ≈70 cm with a focal length of 38 mm and an aperture of f/18, leading to an increased exposure time (Nicolae et al., 2014; Verhoeven, 2016).Six orbits around the object were performed, taking shots with the camera tilted at approximately 45° and 20°, and two sets with frontal shots, ensuring that each point was well visible in a large number of photos (Menna et al., 2017; Collins et al., 2019). To achieve a good image scale, the Ground Sampling Distance (GSD) was calculated to achieve a submillimeter footprint (López-Fernández et al., 2018). This configuration allowed obtaining a set of 281 photos, a significant number given the small size of the object, but useful to ensure high overlap and reconstruction of problematic areas (Zachar et al., 2022).

*5.4 Surveys of the exhibition rooms*

As the primary objective of the project was to recreate the overall experience of the exhibition in the digital realm, it was necessary not only to digitally reconstruct the objects on display but also to faithfully replicate the environment, displays, panels, architecture and spatial layout. The aim was to optimise the dynamic interaction between the objects and their environment, enhancing user understanding and engagement with the digitised exhibition environment.
At the moment, the reconstruction of the spaces is still in progress, and only the first room of the exhibition has been completely acquired in photogrammetry, processed, optimised and modelled in Blender (https://www.blender.org/), as shown in Fig.14. This priority is due to the time constraints previously mentioned, which led the teams to focus on the objects on display first. On the other hand, since the spaces were going to be still accessible after the exhibition's closing, their acquisition was not limited by a fixed timetable, allowing for continuous progress afterwards and flexibility in the work.

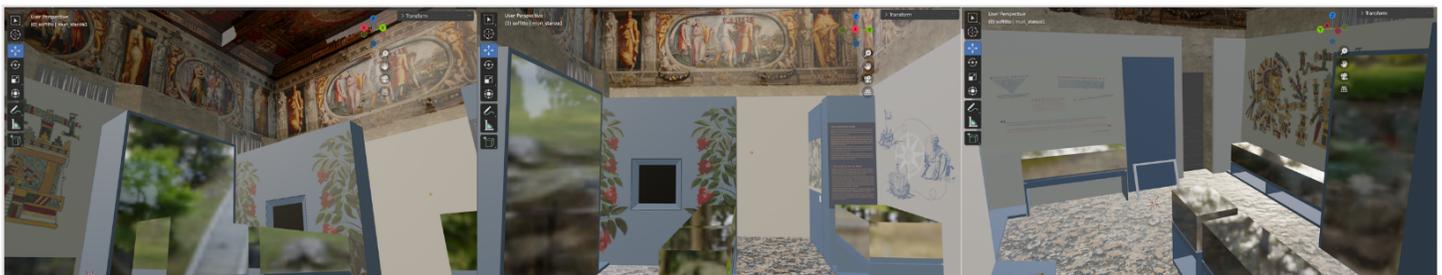
Fig. 14. Modelling process in Blender of the first room of the exhibition at Palazzo Poggi.

While planning the acquisition and modelling of the physical spaces, we had to consider and counterbalance two opposite requirements. On the one hand, we needed to pay attention to the artistic value of Palazzo Poggi's frescoes and paintings on walls and ceilings, some dating back to the 16th century. On the other hand, we needed to obtain a very optimised final model (in terms of geometry and textures) to make it accessible in real-time and online. For this reason, we have adopted a mixed solution that could have considered both aspects. We decided to proceed and acquire only the painted decorations with photogrammetry, keeping most of the details in these areas only, while the rest of the room's architecture, including the displays, the museum cases, etc., was modelled in Blender, starting from the plans/blueprints.

Following this approach, the photogrammetric acquisition was carried out, shooting from the lowest possible angle on the ground due to the high position of the frescoes and friezes. Moreover, special attention was paid to the beams, which were decorated on all sides because of the peculiar structure of the ceiling. Acquired images have been then imported and processed into the SfM software 3DF Zephyr, following a state-of-the-art approach in data processing. Gaps and hidden parts have been then completed directly in Computer Graphics modelling.

To keep track of this double approach, we have created a specific semantic layer that would have identified and specifically kept track of the parts of the models obtained with photogrammetry, distinguishing them from those modelled in Computer Graphics, in line with the principles of London Charter (https://londoncharter.org/).

## 6. Moving to the Virtual Domain

Every project step involves the iterative creation, collection and exchange of complex research data and metadata. Carefully planning data workflows and keeping a detailed record of the processes (individuals and institutions involved, time, place, capture techniques, hardware and software used, etc.) are key for tracking progress and evaluating the project's success. Crucially, they also ensure the quality of the research data, their findability, their accessibility in the long-term, their interoperability and maximum reusability, in line with FAIR principles.

Below, we describe how we have decided to adopt and use the ATON framework to create and deploy the digital twin of Aldrovandi's exhibition (Section 6.1). This discussion is followed by a presentation of how bibliographical data and metadata relating to the digitisation process have been collected and curated and further exploited as Linked Open Data (Section 6.2), how data provenance and subsequent changes have been tracked throughout the project to produce reliable and trustworthy data that are FAIR "by-design" (Section 6.3), and how all these data were extended and visualised for human consumption (Section 6.4).

### *6.1. Towards the digital twin: modelling, optimising and publishing models in ATON Framework*

After the conclusion of the 3D survey and processing, we worked on the optimisation of the geometry and textures of each model. One of the main aims of optimisation is to obtain performing and realistic models suitable for different devices, to ensure interoperability. This phase required automatic and manual approaches to refine visual representation, ensuring the utmost precision and photorealism in the final output.

The starting input was the highest resolution model, obtained after photogrammetry or scanner post-processing (level 0). This model often shows topological problems due to the lack of data and shadow areas during the survey or occluding geometries of the items involved. Main issues include, therefore, holes, non-manifold edges and overlapping faces, depending on the acquisition conditions and item features. For instance, some items, such as manuscripts, could not be handled, due to their delicate state of health, limiting data acquisition to their front and side. In these cases, we have first processed all level 0 models through Instant Meshes, a software that employs a cohesive local smoothing algorithm to re-mesh the surface, enhancing both the edge orientations and vertex positions in the resulting mesh (Jakob et al., 2015). This first processing returned a regular topology. The missing parts have been manually modelled, using references, integrating them and optimising through editing and sculpting tools in Blender. Once the model was fixed in terms of topological issues, a further re-meshing in Instant Mesh was performed to give a regular topology to the reconstructed parts of the mesh and to obtain the final level 1 model (Fig. 15).

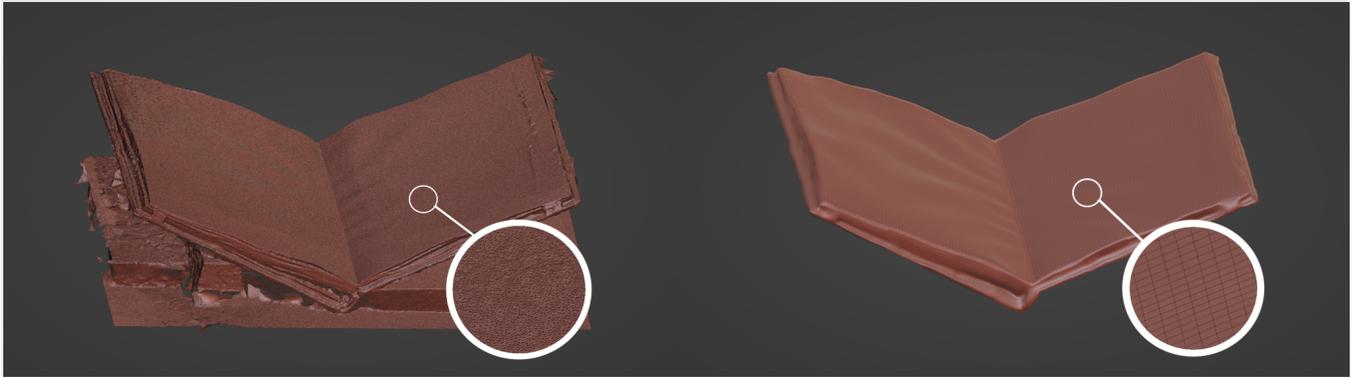

Fig. 15. Example of a geometry reconstruction and optimisation. The images show the Nautical Atlas, an ancient book part of the exhibition, before (level 0) and after the intervention (level 1).

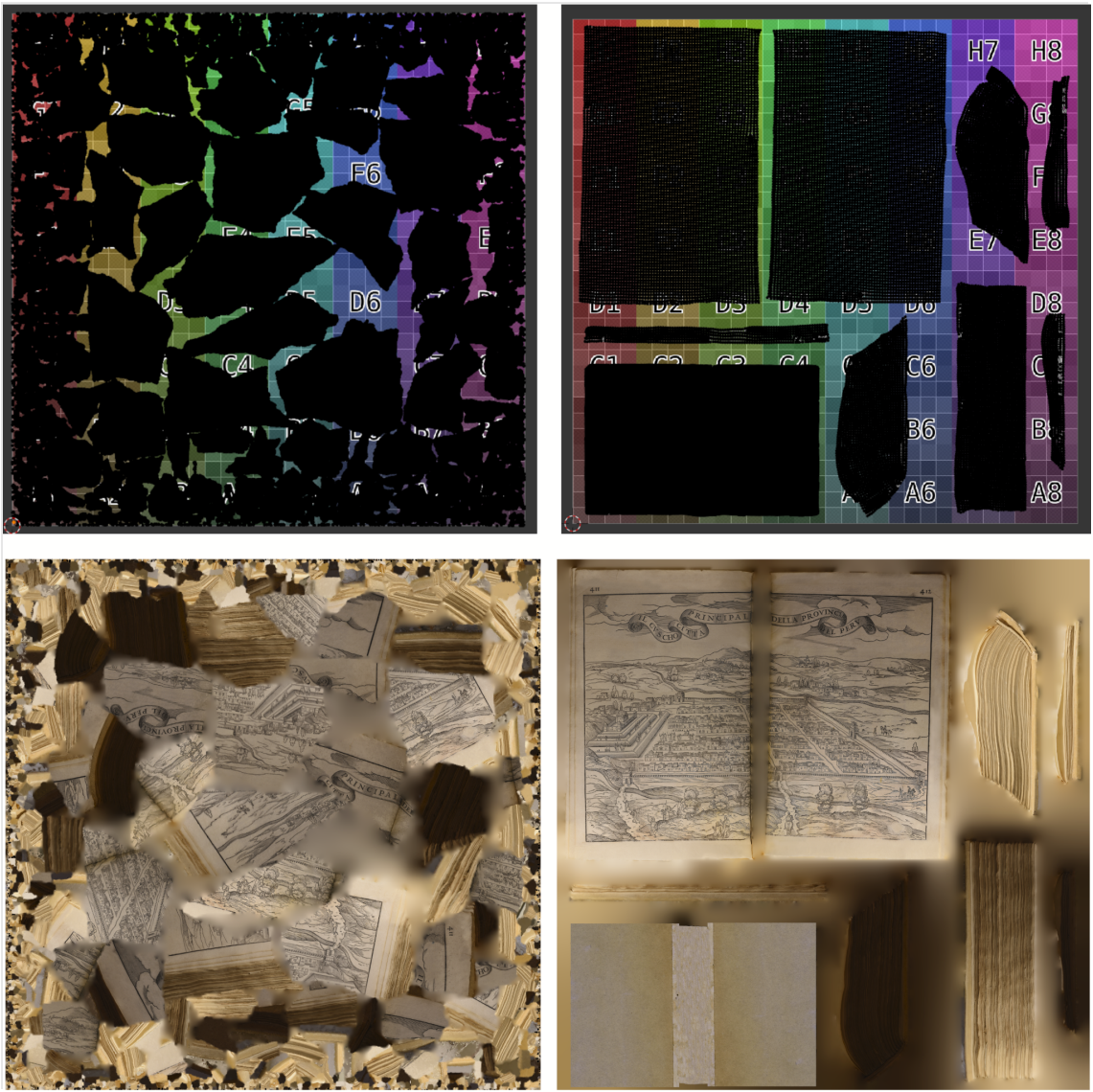

Fig. 16. Difference between UV maps and textures of the Cusco Map before and after the intervention.

The final step required to optimise these models (level 2). Instant Mesh can generate a new low-polygon model while preserving the original shape and boundaries, achieving a pure quad mesh. Then, UV coordinates were created for texture mapping, the procedure with which the software maps 2D images (texture) onto a 3D model surface to simulate colours and achieve a photorealistic appearance. The UV space was organised manually, marking the seam line to define UV islands and control the unwrap. The unwrap is the UV mapping method that allows flattening the mesh surface by cutting along seam lines and fitting the UV island in the UV space, avoiding overlapping.

Depending on each item, the manual process helps to organise the texture space to prioritise the UV island of greatest interest, using as many pixels as possible provided by the UV space. Taking the books as a reference, priority was given to UV islands related to the pages exposed, which take up a relevant space in the UV map.

To perform the texture building and mapping, each model parametrised in Blender has been exported in obj format and re-imported with customised UV in the photogrammetric software previously used to generate them (either 3DF Zephyr or Metashape). Then the atlas textures were generated using software options that allow keeping customised UV in the external software during the computation (shown in Fig. 16).

In some cases, the textures have been manually edited using photo-editing software, such as Gimp (https://www.gimp.org/), or the paint tool of Blender (shown in Fig. 17) to fix small gaps or balance colour and exposure.

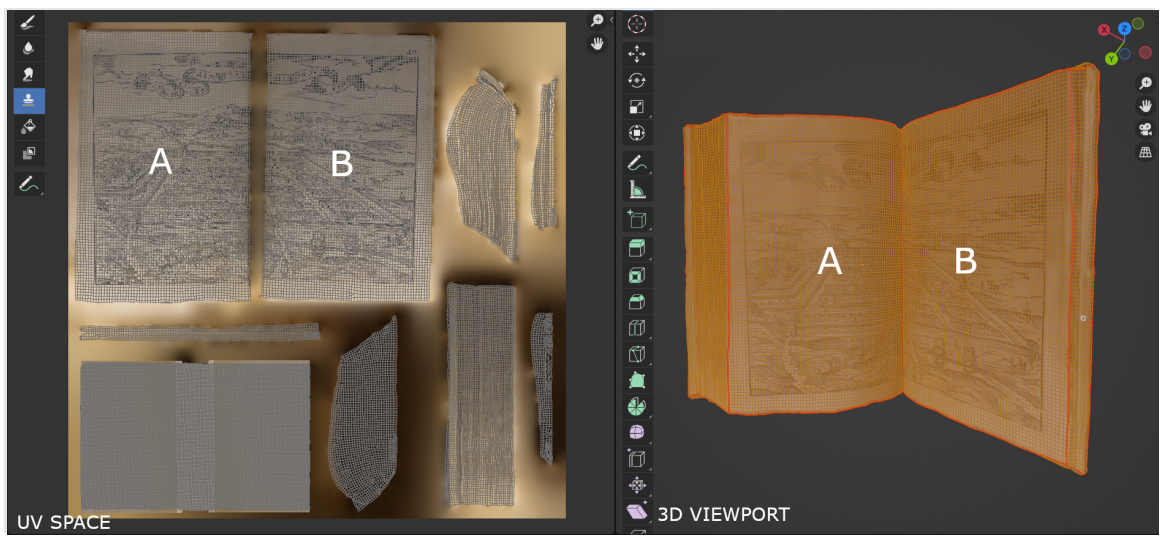

Fig. 17. Texture optimisation after 3D model re-meshing, custom UV mapping, and texture building. The organisation of the UV space (left) shows how the UV islands related to the pages have priority in dimensions (A and B). The image shows the Cusco Map, an ancient book part of the exhibition.

Finally, each model has been exported in glFT format (Robinet et al., 2014), suitable for the navigation and sharing in Web3D applications of each 3D item involved in the exhibition. The export was performed following straightforward guidelines specifically designed for the Web3D platform adopted in this project (https://osiris.itabc.cnr.it/aton/index.php/tutorials/creating-3d-content-for-aton/exporting-3d-models-from-blender/).

Since our goal was to obtain realistic and metadata-enriched models accessible online in 3D and from different devices, we have decided to publish each post-processed model using ATON, the open-source framework developed by the CNR-ISPC (Fanini et al. 2021). Born in 2016, the framework is built on strong web standards and large open-source ecosystems like Three.js (https://threejs.org/) and Node.js (https://nodejs.org/). ATON offers a scalable and adaptable solution for institutions, museums, labs, researchers, and developers who wish to create and deploy cross-device Web3D/WebXR applications tailored for the Cultural Heritage domain (González-Zúñiga and O'Shaughnessy, 2019). The framework also offers built-in client-side and server-side components that enable synchronous collaboration. These features enable the enhancement of online 3D experiences from individual to multi-user settings, opening up exciting possibilities for cultural heritage applications such as virtual forums, educational instruction, team-building activities, and multiplayer 3D gaming (Fanini et al., 2021). With its modular architecture, the framework allows fast development and deployment of custom interactive experiences (web apps) designed around specific requirements without any installation required for end-users (Gonizzi Barsanti et al., 2018; Lo Turco et al., 2019; Fanini et al., 2022).

Within the framework, each 3D scene (a descriptor) can reference one or multiple items in a collection (e.g. a 3D model) particularly suited for 3D galleries within online virtual museums. Each scene is assigned a unique alphanumeric identifier, called *scene ID*, within the currently active instance of the framework. This identifier enables any web app built on ATON to access or operate with specific published 3D scenes. The framework also includes an integrated authentication system that allows content creators and publishers to access, manage, and modify their collections and scenes within the deployed instance. Users can easily place and organise their content in the primary collection folder, which may consist of 3D models, audio files, panoramas, and more. Authors, editors, or content creators can use *Shu*, a user-friendly back-end system, to publish and manage 3D scenes.

## *6.2. Metadata collection, curation, and modelling*

Recreating the Digital Twin required us to collect and organise a variety of information, including (1) bibliographic metadata on the nature of the objects, their provenance, and their subject classification, as well as (2) details on the digitisation process, such as people involved in acquisition activities, methods and tools, and institutional information.

On the one hand, to collect bibliographic metadata, we rely on official sources, such as the official exhibition catalogue, preliminary unstructured notes created by exhibition curators as drafts of the exhibit organisation, and other museum cataloguing records. A team of data curators has reviewed the sources and created a draft list of objects to be described, which roughly includes around 270 records (the actual number of objects is higher, although many similar objects are grouped in complex objects, e.g. panels including 200 similar engravings are grouped into one table row). It is worth noting that, while the original exhibition included videos of artefacts somehow connected to the ones on display, but no further information is available in the catalogue, the final dataset also includes extensive detail of such artefacts to expand the boundaries of the Aldrovandi's collection with meaningful links to heritage objects preserved across museums and cultural institutions, therefore showing the versatile nature and impact of Aldrovandi's studies. On the other hand, the data collection of the digitisation process is performed in a laboratory setting, meaning it is collected while performing the acquisition activities.

Both datasets, i.e. the bibliographic and the digitisation ones, are created using collaborative editing tools and are preliminarily stored in tabular form. Tabular data is easy to understand by all curators regardless of their technical skills, and collaborative editing environments, such as Google Spreadsheet, make it easier to manipulate data in real-time. The definition of tables schemas (i.e. the headings, the definition of expected cell data, and controlled vocabularies of terms to be used in certain columns) has been defined for both tables by data engineers. For instance, controlled vocabularies are defined for types of artefacts and places. Moreover, to ensure consistent reuse of data and to eventually enrich our data with data available in external datasets, we include identifiers defined by well-known bibliographic and cultural authorities (e.g. VIAF by Library of Congress, ULAN by the Getty Research Institute) to unequivocally identify important parts of our data. For instance, bibliographic data include mentions of relevant people (illustrators, painters, clients) and cultural institutions that currently preserve the exhibits. Whenever applicable, people are identified via their VIAF identifier. When available, the Getty ULAN identifier for artists is provided instead. Likewise, institutions are identified by means of a Wikidata identifier and places via their geoNames identifier. The accuracy of data created is evaluated by a pool of curators, who cross-validate metadata created by others.

It is worth noting that tabular data have been created to favour daily updating operations, but it is not meant to be the (only) final format of the datasets. A different approach to publishing the final dataset has been evaluated, that is, Linked Open Data (Bizer et al., 2009). The Linked Open Data approach is particularly suited for publishing cultural heritage data since the technology allows curators to access information available in other catalogues/datasets and to expand their catalogue with pieces of information extracted from such sources. So, we maintain as little data as possible while we preserve the link to official information sources (e.g. artists' biographies in ULAN). In addition, Linked Open Data formats enable data to be machine-readable, machine-interpretable, and machine-actionable (David et al., 2023) and foster the best interoperability, thus being compliant with the FAIR principles of using, as much as possible, a formal, accessible, shared, and broadly applicable language for knowledge representation (Wilkinson et al., 2016).

To exploit the datasets as Linked Open Data, the information structured in tabular format must undergo a crosswalk process to be reshaped into a Resource Description Framework (RDF) format (Cyganiak et al., 2014). Thus, two ad-hoc data models were defined:

(a) the first one is based on the CIDOC Conceptual Reference Model (CIDOC CRM) (Doerr, 2003) and its extension CRM Digital (CRMdig) (Doerr and Theodoridou, 2011), and it is derived from the digitisation dataset to standardise the semantics of the data and promote its interoperability;
(b) the second one is based on CIDOC CRM and it is designed from the bibliographic spreadsheet data to describe the object's characteristics and contextual information.

Because of the advantages of querying data with the Semantic Web technologies, we reused some tools available in the literature to accomplish the non-trivial task of schema crosswalk from the spreadsheet documents into RDF. In particular, we have adopted the RDF Mapping Language (RML) (Dimou et al., 2014) for completing this activity.

### *6.3. Provenance and change-tracking*

In the context of metadata management, provenance and change-tracking are paramount. Provenance provides a record of where data came from, who created it, and when (Gil et al., 2010), while change tracking allows one to understand the whole history of the information associated with an entity, which can change in time due to several situations (extending its description with additional information, modifying some of the data due to possible mistake, etc.). Both provenance and change tracking are essential for ensuring the *reliability* and *trustworthiness* of data.

The provenance model employed in our work is based on the OpenCitations Data Model (OCDM) (Daquino et al., 2020). According to OCDM, a new snapshot is defined every time an entity is created or modified, and it is stored within a (provenance) named graph (Persiani et al., 2022). The snapshots record the validity dates, the agents responsible for the creation/modification of entities' data, the primary sources, a link to the previous snapshot in time, and a human-readable description. Furthermore, OCDM extends the Provenance Ontology (Lebo et al., 2013) by introducing a mechanism to record additions and deletions from an RDF dataset with a SPARQL query (Harris and Seaborne, 2013). This solution allows restoring an entity to a specific snapshot by applying the reverse operations of all update queries from the most recent snapshots to the desired one.

To handle temporal queries, we utilised the *time-agnostic* library (Massari and Peroni, 2022). This Python package enables all six retrieval functionalities considered in the taxonomy by Fernández et al. (2015). The combination of the provenance model, change-tracking model, and the time-agnostic library provides a robust framework for managing metadata, tracking changes, and handling temporal queries in our system.

### *6.4. Data visualisation and data storytelling*

Once all the aforementioned data are publicly available and query able on the Web using Linked Open Data technologies, cataloguing records are presented in an online digital library, which includes all data recorded in the aforementioned tables and additional information extracted from external sources referenced in our data. For instance, geo data are extracted from Geonames (https://www.geonames.org/) to geolocalise the exhibits in maps. Similarly, artists and their floruit are extracted from ULAN and VIAF, to understand the time range of objects displayed in the exhibition, as well as artistic movements relevant to the artists.

To quantitatively analyse and visualise aggregated data, we use MELODY, an online dashboarding system for creating web-ready data stories that leverage Linked Open Data (Daquino and Renda, 2023). In particular, MELODY allows users to explore one or more Linked Open Data sources via SPARQL queries. Query results are attached to User Interface components (e.g. charts, maps, graphs, text searches, tables) which can be selected, added, moved, and displayed in a canvas according to the data story creator, who can also alternate components with curated text, therefore contextualising charts and providing their interpretation of data. The final data story, as well as each UI component, can be exported and embedded in other web pages, therefore enriching the catalogue of Aldrovandi's exhibition with figures on relevant, quantitative insights. Data storytelling options are available to both data curators, who can select appropriate (dynamic) visualisations to be included in the digital library, and to end users, who can explore Aldrovandi's dataset via MELODY online platform (https://projects.dharc.unibo.it/melody/) and publish their own data stories in a dedicated catalogue (https://melody-data.github.io/stories/).

Lastly, the 3D models of digitised objects are included in digital library records, which are provided by ATON. In turn, metadata can be openly accessed and queried by external applications, such as the aforementioned Aton, that include details on demand while exploring the exhibition.

## 7. Discussions and Conclusion

In the previous sections, we have shown our approach to enabling the preservation and accessibility of exhibitions (RQ1). Our solution is entirely based on adopting digitisation techniques and tools that permit the creation of a digital representation of a real event with all the objects and narratives it defined and to make such a *digital twin* available by using appropriate Web technologies. We reached that goal first by adopting a hybrid use of two distinct acquisition techniques, i.e. structured light projection scanning and photogrammetry, which enabled us to create new digital cultural heritage objects and environments derived from those available in the rooms of Palazzo Poggi at the University of Bologna, where the temporary exhibition entitled "The Other Renaissance: Ulisse Aldrovandi and the wonders of the world" was originally set. Second, we stressed the adoption of as-open-as-possible technologies (e.g. ATON), formats (e.g. open standards for digital 3D images) and protocols (e.g. Web-based standards) to make available the final digital product to the users.

We have investigated the existing literature to assess possible constraints that would have characterised the acquisition process – i.e. time, space, and materials – that we experienced during the work (RQ2). The first factor, i.e. time, was a direct consequence of the temporary nature of an exhibition. Indeed, we had to work on an *ongoing* temporary event, with a limited time available dictated by the fact that the exhibition was open the whole week but Monday, which gave us, on average, only one day per week to remove the display cases to do the acquisition processes and to re-install them for preparing the exhibition to be visited the day after. In addition, some objects did not belong to the permanent collection of Palazzo Poggi and had to be returned to the reality they belonged at the end of the exhibition. Another factor making the work more complex was, indeed, related to space constraints. Indeed, it was challenging to organise the acquisition activities involving all the research groups in Palazzo Poggi due to its limited available space to move the display cases and to bring in all the technical equipment needed for the activity. All these activities, indeed, were possible thanks to the incredible in-kind effort of the staff of the University of Bologna Museum Network, who also mediated with all the other cultural institutions involved – the Bologna University Library, the Medieval Civic Museum of Bologna, the Archaeological Civic Museum and the State Archives of Bologna, the Carrara Academy of Bergamo, the Museum of Civilisations and the Spada Gallery of the city of Rome, the Academy of Physiocritics of Siena, and finally the Natural History Museum of Verona.

An additional problem related to the acquisition concerned the materials of which some works are composed (RQ3). Due to the heterogeneity of the objects included in the temporary exhibition, we addressed the huge challenge of acquiring objects having different, non-cooperative or hybrid materials, including specular components that influenced the optical response of the detection instruments (e.g. black, glossy and transparent surfaces). Section 5.3 detailed some of the most peculiar objects and the approaches we have adopted to bypass or mitigate the issues.

As far as RQ4 is concerned, we have described in detail our research data workflow in Section 6. Bibliographical data and metadata relating to the acquisition process have been collected and curated to be further exploited as Linked Open Data, analysed, visualised and enriched with external data sources. Many specific formats, schemas and vocabularies, as well as identifiers defined by well-known bibliographic and cultural authorities, have been employed from the first stages of data production. Then, we identified the need to convert this data from an initial tabular format (preferred for its ease of use across a sizable project team) into a Resource Description Framework format via a crosswalk based on the CIDOC-CRM. This allows the presentation of cataloguing records in an online digital library where they can be enriched with information extracted from external sources (e.g. geographical data), quantitatively analysed, visualised in an aggregated and curated form, and queried by external applications such as ATON, that is the framework we chose to allow the fast development and deployment of custom interactive experiences via the Web, thus without requiring any software installation on the part of end-users.

Finally, to respond to RQ5, the approach proposed in this paper certainly implies the availability of equipment and experts proper to a research laboratory or specialized companies/realities for the acquisition process. In fact, the equipment used here, such as scanners and professional cameras, requires good economic availability and trained and qualified personnel who can make the best use of each of their features and find effective solutions in case of unforeseen events – such as those introduced in Section 5.3.

Often, a museum or cultural institution organising a temporary exhibition lack the budget to purchase/borrow and exploit such equipment and personnel. One solution would be, as in our case, to take advantage of an agreement with university research laboratories as part of national and international projects, internships, and master students' work for their thesis focused on digitising cultural heritage. In this way the available budget of the museums could be redirected to extra payments, such as those related to the extraordinary opening of display cases.

However, if a museum cannot plan the acquisition and digitisation project as above, there are other affordable (in terms of costs and human expertise) acquisition tools that can be adopted for the digital reproduction of objects and exhibition spaces. This is the case of free photogrammetry apps available on the App Store or Google Play – like RealityScan, Qlone, or Scann3D – downloadable from smartphones and usable even by non-expert users for digitising objects, even if the results are not comparable with those obtained by a digitisation process such as the one described in the present article. In addition, for the virtual representation of exhibition spaces, an economic alternative would be the creation of a Virtual Tour. Indeed, during our work, we have used this technology for creating an in-house documentation for keeping track of the structure of the exhibition, as introduced in Section 4.3. For implementing Virtual Tours, one would only require a 360-degree camera (some are available at a reasonable price) for obtaining panoramic images of the spaces and assembling these into a real interactive tour via free apps such as 3DVista (https://www.3dvista.com/en/apps).

Summarising, the workflow we have introduced in this article seems to provide a robust solution for managing data and maximising their findability, accessibility, interoperability and reusability in other contexts. It goes well beyond the minimum requirements in terms of FAIRness of the generated data: it combines the use of standard and open file formats, open tools and software (wherever possible), schemas, vocabularies and identifiers with a system to convert and enrich them as Linked Open Data, all the while tracking their provenance and subsequent changes.

Of course, challenges remain in connecting 3D and multimedia data with their metadata and spatial location in the exhibition. The experimentation regarding 3D cultural heritage models is ongoing, and some aspects, especially regarding data management according to FAIR principles, are still to be explored in detail (Knazook et al., 2023a,b). However, initiatives such as the Italian National Plan for the Digitisation of Cultural Heritage and the relative guidelines on the creation, metadata processing and archiving of digital objects will certainly begin to close this gap in the near future.

Moreover, at this stage, we have focused on creating the Digital Twin, while more complex aspects connected to the re-use of the Twin for advanced XR purposes are ongoing. Specifically, one of our initial goals was to recreate the experience of the exhibition for the visitors, and this requires storytelling (i.e. automatic guiding audio narrative), social participation (i.e. co-presence and collaboration in the 3D immersive space), authentic guided tours in this no more existing exhibition, organised by the institution itself or by experts involving groups of visitors or families from home (Pescarin et al., 2023b; Pescarin et al., 2023c). In addition, we aim also at developing a prototype for a full virtual reality implementation of the exhibition, to be consumed by users with appropriate hardware – e.g. virtual reality (VR) headsets. All these scenarios will be experimented in future activities of our research. An aspect that deserves special attention is linked to the long-term preservation of the research data produced in a project of this kind. In particular:

- How do we reconcile the long-term data preservation in a suitable infrastructure with the needs of a virtual exhibition?
- In a scenario where researchers collaborate with many cultural institutions to digitise their holdings, who is responsible for the long-term preservation of the data produced?
- Does everything need to be preserved long-term? And, if not, who decides what data to keep and how?
- What infrastructures are available for the preservation of 3D cultural heritage data? And what can the experience of museums that maintain their research data repositories teach us?

Creating the digital twin of Aldrovani's exhibition is still ongoing while we are writing. While the acquisition and processing of the objects is completed, we are currently working on the last phases concerning the publication of the digital representation of all the objects in the exhibition and, finally, the release of the final digital twin of the experience. As an immediate next step, we will soon make available the digital representation of the first room of the exhibition, with all its objects. The remaining rooms will be made available during the next months. In addition, another aspect we will work on in the next months concerns the release of a knowledge graph, which would provide a detailed record of the exhibition, including information on the digitised objects, their characteristics, the digitisation process, and the individuals and institutions involved. Such a knowledge graph and digital twins of both the exhibition and the single objects could be a valuable resource for researchers and cultural heritage professionals and a flexible framework that could be adapted to meet the specific needs of different stakeholders in the cultural heritage domain.

Regarding the Project CHANGES, this work has allowed us to improve the understanding of planning, acquiring, processing, and online publication tasks, thus enabling us to scale them up to wider case studies and to predict the time, technology, personnel, and resources needed in the *core* case studies of the project.

**Acknowledgements**

The authors want to thank the personnel of the cultural institutions involved in this project, i.e. the University of Bologna Museum Network (Annalisa Managlia, Cristina Nisi, Roberta Trini, Silvia Matteucci), the Bologna University Library (Giacomo Nerozzi, Stefania Filippi), the Archeological Museum of Bologna (Daniela Picchi), and the Medieval Civic Museum of Bologna (Ilaria Negretti). In addition, the authors want to thank the students of the Master's Degree in Digital Humanities and Digital Knowledge (Daniele Spedicati, Marco Lamorte, Matteo Guenci, Stefano Sorrentino), the Master's Degree in Engineering Of Building Processes and Systems (Dorsa Rashidpour, Elnaz Bostani, Ghazaleh Farzian) and the Master's Degree in Science for the Conservation-restoration of Cultural Heritage (Ashraquet Alphonse Fakhry Bastawrous). This work has been partially funded by the European Union's NextGenerationEU economic recovery package supporting the Project CHANGES - Cultural Heritage Active Innovation For Next-Gen Sustainable Society (PE 0000020, CUP B53C22003780006).

Kong, X., Hucks, R.G., 2023. Preserving our heritage: A photogrammetry-based digital twin framework for monitoring deteriorations of historic structures. Automation in Construction 152, 104928. https://doi.org/10.1016/j.autcon.2023.104928

Koster, L., Woutersen-Windhouwer, S., 2018. FAIR Principles for Library, Archive and Museum Collections: A proposal for standards for reusable collections. The Code4Lib Journal.

Kraus, K., 2007. Photogrammetry: Geometry from Images and Laser Scans. De Gruyter.

Kritzinger, W., Karner, M., Traar, G., Henjes, J., Sihn, W., 2018. Digital Twin in manufacturing: A categorical literature review and classification. IFAC-PapersOnLine 51, 1016–1022. https://doi.org/10.1016/j.ifacol.2018.08.474

La Russa, F.M., Santagati, C., 2020. Historical Sentient - Building Information Model: a Digital Twin for the Management of Museum Collections in Historical Architectures. Int. Arch. Photogramm. Remote Sens. Spatial Inf. Sci. XLIII-B4-2020, 755–762. https://doi.org/10.5194/isprs-archives-XLIII-B4-2020-755-2020

Latini, E., 2005. Museo di Palazzo Poggi, in: Bologna. Una Provincia, Cento Musei. Pendragon, Bologna.

Lebo, T., Sahoo, S., & McGuinness, D., 2013. 04 30). PROV-O: The PROV Ontology. http://www.w3.org/TR/2013/REC-prov-o-20130430/

Lo Brutto, M., Spera, M.G., 2011. Sperimentazione di procedure automatiche in fotogrammetria close range per il rilievo di Beni Culturali, in: Atti Della 15a Conferenza Nazionale Delle Associazioni Scientifiche per Le Informazioni Territoriali e Ambientali (A.S.I.T.A.). Reggia di Colorno, pp. 1427–1438.

Lo Turco, M., Piumatti, P., Calvano, M., Giovannini, E.C., Mafrici, N., Tomalini, A., Fanini, B., 2019. Interactive Digital Environments for Cultural Heritage and Museums. Building a digital ecosystem to display hidden collections. https://doi.org/10.20365/DISEGNARECON.23.2019.7

López, F.J., Lerones, P.M., Llamas, J., Gómez-García-Bermejo, J., Zalama, E., 2018. A Review of Heritage Building Information Modeling (H-BIM). Multimodal Technologies and Interaction 2, 21. https://doi.org/10.3390/mti2020021

López-Fernández, L., Lagüela, S., Rodríguez-Gonzálvez, P., Martín-Jiménez, J., González-Aguilera, D., 2018. Close-Range Photogrammetry and Infrared Imaging for Non-Invasive Honeybee Hive Population Assessment. IJGI 7, 350. https://doi.org/10.3390/ijgi7090350

Luhmann, T., Robson, S., Kyle, S., Boehn, J., 2020. Close range photogrammetry and 3D imaging. Whittles Publishing, Caithness, Scotland.

Marziali, S., Dionisio, G., 2017. Photogrammetry and macro photography. The experience of the MUSINT II Project in the 3D digitizing process of small size archaeological artifacts. Studies in Digital Heritage 1, 298–309. https://doi.org/10.14434/sdh.v1i2.23250

Massari, A., Peroni, S., 2022. Performing live time-traversal queries via SPARQL on RDF datasets. https://doi.org/10.48550/ARXIV.2210.02534

McPherron, S.P., Gernat, T., Hublin, J.-J., 2009. Structured light scanning for high-resolution documentation of in situ archaeological finds. Journal of Archaeological Science 36, 19–24. https://doi.org/10.1016/j.jas.2008.06.028
39

## Author Contributions

Balzani Roberto: *Writing – original draft, Writing – review & editing*

Barzaghi Sebastian: *Data curation, Investigation, Methodology, Writing – original draft, Writing – review & editing*

Bitelli Gabriele: *Methodology, Resources, Writing – review & editing*

Bonifazi Federica: *Investigation, Methodology, Visualization, Writing – original draft*

Bordignon Alice: *Investigation, Methodology, Visualization, Writing – original draft*

Cipriani Luca: *Methodology, Resources, Writing – original draft, Writing – review & editing*

Colitti Simona: *Writing – review & editing*

Collina Federica: *Methodology, Software, Writing – original draft, Writing – review & editing*

Daquino Marilena: *Data curation, Investigation, Writing – original draft*

Fabbri Francesca: *Writing – original draft, Writing – review & editing*

Fanini Bruno: *Software, Writing – review & editing*

Fantini Filippo: *Methodology, Resources, Writing – original draft, Writing – review & editing*

Ferdani Daniele: *Investigation, Methodology, Resources, Visualization, Writing – original draft*

Fiorini Giulia: *Methodology, Software, Writing – original draft, Writing – review & editing*

Formia Elena: *Writing – review & editing*

Forte Anna: *Methodology, Software, Writing – original draft, Writing – review & editing*

Giacomini Federica: *Methodology, Software, Writing – original draft, Writing – review & editing*

Girelli Valentina Alena: *Methodology, Software, Writing – review & editing*

Gualandi Bianca: *Writing – original draft, Writing – review & editing*

Heibi Ivan: *Data curation, Formal Analysis, Investigation, Resources, Software*

Iannucci Alessandro: *Writing – review & editing*

Manganelli Del Fà Rachele: *Investigation, Methodology, Writing – original draft*

Massari Arcangelo: *Data curation, Software, Writing – original draft, Writing – review & editing*

Moretti Arianna: *Data curation, Formal Analysis, Investigation, Software, Writing – original draft*

Peroni Silvio: *Conceptualization, Data curation, Funding acquisition, Methodology, Project administration, Resources, Supervision, Writing - original draft, Writing - review & editing*

Pescarin Sofia: *Conceptualisation, Methodology, Resources, Supervision, Writing – original draft, Writing – review & editing*

Renda Giulia: *Data curation, Investigation, Resources, Software, Visualization*

Ronchi Diego: *Data curation*

Sullini Mattia: *Methodology, Resources, Writing – original draft, Writing – review & editing*

Tini Maria Alessandra: *Methodology, Software*

Tomasi Francesca: *Writing – review & editing*

Travaglini Laura: *Methodology, Investigation, Visualization, Writing – original draft*

Vittuari Luca: *Resources, Writing - review & editing*